%% file: BDX-PAC46-upd.tex
\documentclass[letterpaper]{article}
\pdfoutput=1 
\usepackage{color,amsmath}
\usepackage{verbatim}
\usepackage{multirow} 
\usepackage{draftwatermark}
\SetWatermarkLightness{1.0} 
\SetWatermarkScale{0.005} 

\newif\ifhyprf   
\hyprftrue
    
\RequirePackage{ifpdf}

\ifpdf
    \pdfoutput=1        
    \pdftrue
\else
    \pdffalse           
\fi

\ifpdf
  \definecolor{rltred}{rgb}{0.75,0,0}
  \definecolor{rltgreen}{rgb}{0,0.3,0}
  \definecolor{rltblue}{rgb}{0,0,0.75}
  \definecolor{rltdarkgreen}{rgb}{0.1,0.6,0.1}
  \ifhyprf
     \usepackage[pdftex,
         colorlinks=true,
         urlcolor=rltblue,       
         filecolor=rltgreen,     
         linkcolor=rltred,       
         citecolor=rltdarkgreen, 
         pagebackref,
         pdfpagemode=None,
         pdftitle={Dark matter search in a Beam-Dump eXperiment (BDX) at Jefferson Lab: an update on PR12-16-001},
         pdfauthor={many},
         pdfsubject={Beam dump experiment JLab Hall A},
         pdfkeywords={Hall A, Beam dump, dark photon}]{hyperref}
  \fi
  \usepackage{pdfcolmk}
  \DeclareGraphicsExtensions{.pdf,.png,.jpg}
\else
  \usepackage{graphicx}
  \DeclareGraphicsExtensions{.eps,.epsi,.ps,.eps.gz,.epsi.gz,.ps.gz}
\fi

\usepackage{graphicx,lscape,rotating}
\usepackage{amsmath}
\usepackage{amssymb,amsfonts,textcomp}

\usepackage{lineno}

\input{newcomm.tex}

\usepackage{alphalph}
\makeatletter
\newcommand*{\fnsymbolsingle}[1]{%
\ensuremath{%
\ifcase#1%
\or *%
\or \dagger
\or \ddagger
\or \mathsection
\or \mathparagraph
\else
\@ctrerr \fi
}%
}
\makeatother
\newalphalph{\fnsymbolmult}[mult]{\fnsymbolsingle}{}

\begin{document}

\begin{center}
{\tiny \leftline{V2.0}}
{\tiny\leftline{\thedate}}
\date{\today}
\rightline{PR12-16-001 update to PAC 46}
\vskip 1.0cm
{\bf\huge Dark matter search in a Beam-Dump eXperiment (BDX) at Jefferson Lab
\\ -- 2018 update to PR12-16-001}

\vskip 0.5cm
{  \large \it The BDX Collaboration }

\vskip 0.5cm
{M.~Battaglieri\footnote{Contact Person, email: Marco.Battaglieri@ge.infn.it}\footnote{Spokesperson}, A.~Bersani, G.~Bracco, B.~Caiffi, A.~Celentano$^\dag$, R.~De~Vita$^\dag$,   L.~Marsicano, P.~Musico, F.~Panza, M.~Ripani
% ...
\\}
{\it\small\msu}
\bigskip 

{K.~Hicks\\}
{\it\small\ohio}
\bigskip

{P.~Cole\\}
{\it\small\lamar}
\bigskip

{S.~Dobbs\\}
{\it\small\fsu}
\bigskip

{ C.~Fanelli, P.~Mohanmurthy\\}
{\it\small\mito}
\bigskip

\newpage
\begin{abstract}

\textcolor{red} {}
This document complements and completes what was submitted last year to PAC45~\cite{bdx-update-PAC45} as an  update to the proposal {\it PR12-16-001 Dark matter search in a Beam-Dump eXperiment (BDX) at Jefferson Lab}~\cite{bdx-proposal}  submitted to JLab-PAC44 in 2016.
Following the suggestions contained in the PAC45 report, in coordination with the lab,  we ran a test to assess the beam-related backgrounds and validate the simulation framework used to design the BDX experiment. Using a common Monte Carlo framework for the test and the proposed experiment, we optimized the selection cuts to maximize the reach considering simultaneously the signal, cosmic-ray background (assessed in Catania test with BDX-Proto) and beam-related backgrounds (irreducible NC and CC neutrino interactions as determined by simulation). Our results confirmed what was presented in the original proposal: with 285 days of a parasitic run at 65$\mu$A  (corresponding to 10$^{22}$ EOT) the BDX experiment will lower the exclusion limits in the case of no signal by one to two orders of magnitude in the parameter space of dark-matter coupling versus mass.

\end{abstract}

\vskip 1.0cm
 
\end{center} 

\newpage
\tableofcontents
\newpage


\input{BDX-PAC46-upd-intro}

\input{BDX-PAC46-upd-th}

\input{BDX-PAC46-mutest}
\input{BDX-PAC46-opt}
\input{BDX-PAC46-opt2}

\section{Summary}
In this update we report on the tests performed in spring 2018 to assess the beam-on background for the BDX experiment. We measured the fluence of muons produced by interactions of the 10.6 GeV electron beam in the Hall-A beam-dump and propagating to the location of the future BDX detector. The measurements were compared to detailed FLUKA and GEANT4 simulations. The good agreement between the measured and expected rates demonstrates we have a good control of the simulation framework used to optimize the BDX experimental set-up. As suggested by PAC45, the same simulation framework was used to study the CC and NC neutrino interactions, which represent irreducible backgrounds of the BDX experiment. We optimized the selection cuts to obtain the maximum reach considering simultaneously the signal, the cosmic background (as measured by tests in Catania) and  the neutrino background (as predicted by simulation).

\clearpage
\clearpage

\appendix
\input{BDX-PAC46-upd-appxA}

\newpage
\clearpage
\input{BDX-PAC46-upd-appxB}
\newpage
\clearpage
\input{BDX-PAC46-upd-appxC}

\clearpage
\clearpage

\newpage

\bibliographystyle{unsrt}  
\bibliography{BDX-PAC46-bib}

\end{document}

%% file: newcomm.tex
\newcommand{\be}{\begin{eqnarray}}
\newcommand{\ee}{\end{eqnarray}}
\def\lsim{\mathrel{\rlap{\lower4pt\hbox{\hskip 0.5 pt$\sim$}}
\raise1pt\hbox{$<$}}}                
\def\gsim{\mathrel{\rlap{\lower4pt\hbox{\hskip1pt$\sim$}}
\raise1pt\hbox{$>$}}}

\newcommand{\apr}{{A^\prime}}

\makeatletter
\newcommand\arraybslash{\let\\\@arraycr}
\makeatother
\newcommand{\thedate}{\today}

\def\gevc2{(GeV/c)$^2$}
\newcommand*{\jlab}{Jefferson Lab, Newport News, VA 23606, USA}
\newcommand*{\nhs}{University of New Hampshire, Durham NH 03824, USA}

\newcommand*{\frascati}{Istituto Nazionale di Fisica Nucleare, Laboratori Nazionali di Frascati, P.O. 13, 00044 Frascati, Italy}
\newcommand*{\genova}{Istituto Nazionale di Fisica Nucleare, Sezione di Genova e Dipartimento di Fisica dell'Universit\`a, 16146 Genova, Italy}

\newcommand*{\lasapienza}{Istituto Nazionale di Fisica Nucleare, Sezione di Roma e  Universit\`a La Sapienza, 00185 Roma, Italy}
\newcommand*{\infnct}{Istituto Nazionale di Fisica Nucleare, Sezione di Catania e Dipartimento di Fisica dell'Universit\`a, Catania, Italy}
\newcommand*{\infnbo}{Istituto Nazionale di Fisica Nucleare, Sezione di Bologna e Dipartimento di Fisica dell'Universit\`a, 40127 Bologna, Italy}
\newcommand*{\infnba}{Istituto Nazionale di Fisica Nucleare, Sezione di Bari e Dipartimento di Fisica dell'Universit\`a, Bari, Italy}
\newcommand*{\infnfe}{Istituto Nazionale di Fisica Nucleare, Sezione di Ferrara e Dipartimento di Fisica dell'Universit\`a, Ferrara, Italy}
\newcommand*{\infnle}{Istituto Nazionale di Fisica Nucleare, Sezione di Lecce 73100 Lecce, Italy}

\newcommand*{\ipn}{Institut de Physique Nucleaire d'Orsay, IN2P3, BP 1, 91406 Orsay, France}

\newcommand*{\lamar}{Lamar University, Department of Physics, Pocatello, Beaumont, Texas
77710 }

\newcommand*{\ohio}{Ohio University, Department of Physics, Athens, OH 45701, USA}
\newcommand*{\gwu} {The George Washington University, Washington, D.C., 20052}

\newcommand*{\odu}{Old Dominion University, Department of Physics,
Norfolk VA 23529, USA}

\newcommand*{\edinb}{Edinburgh University, Edinburgh EH9 3JZ, United Kingdom}
\newcommand*{\glasgow}{University of Glasgow, Glasgow G12 8QQ, United Kingdom}

\newcommand*{\fsu}{Florida State University, Tallahassee, , Florida 32306}

\newcommand{\sassari}{Universit\`a di Sassari e Istituto Nazionale di Fisica Nucleare, 07100 Sassari, Italy}
\newcommand{\torvergata} {Istituto Nazionale di Fisica Nucleare, Sezione di Roma-TorVergata e Dipartimento di Fisica dell'Universit\`a, Roma, Italy}

\newcommand{\torino} {Istituto Nazionale di Fisica Nucleare, Sezione di Torino,  Torino, Italy}
\newcommand{\padova} {Istituto Nazionale di Fisica Nucleare, Sezione di Padova,  Padova, Italy}
\newcommand*{\hu}{Department of Physics, Hampton University, Hampton VA 23668, USA }
\newcommand*{\canisius}{Canisius College, Buffalo NY 14208, USA}
\newcommand*{\occidental}{Occidental College, Los Angeles, California 90041, USA}
\newcommand*{\unm}{University of New Mexico, Albuquerque, New Mexico, NM}

\newcommand*{\slac}{Stanford Linear Accelerator Center (SLAC), Menlo Park, CA 94025, US}
\newcommand*{\fnal}{Center for Particle Astrophysics, Fermi National Accelerator Laboratory, Batavia, IL 60510}
\newcommand*{\msu}{Mississippi State University, Mississippi State, MS 39762, USA}
\newcommand*{\stony}{C.N. Yang Inst. for Theoretical Physics, Stony Brook University, NY}

\newcommand*{\spaolo}{Instituto de Fisica, Universidade de S\~ao Paulo, Brasil}

\newcommand*{\julich}{Nuclear Physics Institute and Juelich Center for Hadron Physics, Forschungszentrum Juelich, Germany}
\newcommand*{\mito}{Massachusetts Institute of Technology, Cambridge, MA 02139, USA}
\newcommand*{\bnl}{Brookhaven National Laboratory, Upton, NY 11973, USA}

%% file: BDX-PAC46-upd-intro.tex
\section{Executive summary}
\label{sec:intro}
This document complements and completes the update to 
 the proposal {\it PR12-16-001 Dark matter search in a Beam-Dump eXperiment (BDX) at Jefferson Lab}~\cite{bdx-proposal} presented last year~\cite{bdx-update-PAC45}. 
 
We report here on the test presented to PAC45 and performed in spring 2018  to assess the beam-on background produced by the high current ($\sim$ 20 $\mu$A), high energy  (10.6 GeV) electron beam impinging on  the Hall-A beam-dump. 
Using a detector (BDX-Hodo) that shares the same technology proposed for BDX (CsI(Tl) crystal + plastic scintillator and WLS fiber with SiPM reaodut) we measured the muon fluence at the location of the future BDX detector.  In the current unshielded configuration, muons propagate in the dirt up to the detector location.   The measurement, performed at 25.7 and 28.8 m downstream of the beam-dump, provided the absolute muon rate at beam height and at different vertical positions up to 1 m away from the beam-line center.
The agreement in absolute value and shape between data and simulation demonstrates that the simulation framework based on FLUKA and GEANT4 we used to estimate  the BDX beam-related  background is reliable. 

As predicted by simulation, no significant effect (pile-up or single-hit) of high energy neutrons (T$_N>$10 MeV) has been observed in the crystal. Some runs were also taken with a lower beam energy (4.3 GeV), in a condition   where all  muons are expected to range-out. These showed that for an accumulated charge corresponding to 10$^{20}$ EOT (1$\%$ of full BDX EOT), 
the high energy part of the crystal  spectrum is compatible with cosmic-ray background only. 
With the validation of the simulation framework, we confirm that neutrinos are the only source of  beam-related background since the other particles  are either ranged-out by the planned shielding (muons, gamma and electrons) or do not deposit sufficient energy into the BDX detector to pass our selection cuts (neutrons).  The results of these detailed studies validate the background expectations presented in the original proposal. 

The same simulation framework was used to optimize the BDX experimental setup by maximizing the expected reach. The best configuration was determined by evaluating the sensitivity of the experiment to various selection criteria. Selection cuts were applied consistently on the signal (the simulated electromagnetic shower induced by the $\chi$-electron interaction), the cosmic-ray background (as measured with the BDX-Prototype in Catania and LNS) and the irreducible beam-related backgrounds (charged current (CC) and neutral current (NC) neutrino interactions in BDX detector) and evaluated by their effect on interesting regions of the parameter space of dark-matter coupling versus mass. 
 
This document is organised as follows: a brief theoretical update as well as a brief discussion about the complementarity of BDX  with-respect-to other experiments  is provided in Sec.~\ref{sec:th}; 
results of tests carried out at JLab in spring 2018 are reported in Sec.~\ref{sec:mutest}; Montecarlo simulations of beam-related background expected in BDX  and the reach optimization are reported in Sec.~\ref{sec:opt}.

%% file: BDX-PAC46-upd-th.tex
\section{Theory update}
\label{sec:th}

There have been no substantial changes to the status of light dark matter searches since our update to PAC45. However, the ``US Cosmic Visions: New Ideas in Dark Matter 2017 : Community Report''\,\cite{Battaglieri:2017CVR} was issued and contains a comprehensive overview of the field. In this section we first review the theoretical underpinning of the program (Section\,\ref{sec:theory}), including some updated applications of the report to BDX. We then highlight the strengths of BDX relative to other experimental efforts (Section\,\ref{sec:complementarity}).

\subsection{Review of Light Thermal Dark Matter}
\label{sec:theory}
In this section we review representative models of sub-GeV Dark Matter (DM) as presented more comprehensively in Refs.~\cite{bdx-proposal}.
If the dark and visible matter have sufficiently large interactions to achieve thermal equilibrium during the early universe, the resulting DM abundance greatly exceeds the observed density in the universe today; thus, a  thermal origin requires a sufficient DM annihilation rate to deplete this excess abundance and agree with observation at later times. For thermal dark matter below the GeV scale, this requirement can only be satisfied if the dark sector contains comparably light new force carriers to mediate the necessary annihilation process. 
Such ``mediators" must couple to visible matter and be neutral under the Standard Model (SM) gauge group, so the options for possible mediators can be enumerated in an economical list.   

\subsubsection{Dark Photon Mediator} 
\label{sec:dark-photon}
A popular representative model involves a so-called ``dark photon" $\apr$ with mass $m_{\apr}$ and Lagrangian in the interaction basis
\cite{Holdom:1985ag}
 \be
 \label{eq:lagrangian}
{\cal L} =
-\frac{1}{4}F^\prime_{\mu\nu} F^{\prime\,\mu\nu} + \frac{\epsilon}{2} F^\prime_{\mu\nu} F_{\mu \nu} + \frac{m^2_{A^\prime}}{2} A^{\prime}_\mu A^{\prime\, \mu} + g_D \apr_\mu J^\mu_D    ,~~ ~~~~~~~~(\rm Interaction~ Basis)
\ee
where  $F^\prime_{\mu\nu} \equiv \partial_\mu A^\prime_\nu -  \partial_\nu A^\prime_\mu$ is the dark photon field strength,
$F_{\mu\nu} \equiv \partial_\mu A_\nu -  \partial_\nu A_\mu$ is the electromagnetic field strength,
  $g_D \equiv \sqrt{4\pi \alpha_D}$ is the dark gauge coupling,  $J^\mu_D$ is the current of DM fields, and $\epsilon$ parametrizes the degree of kinetic mixing between dark and visible photons. 
Although the interaction basis Lagrangian initially has no coupling between the $A^\prime$ and SM particles, 
diagonalizing the kinetic term in Eq.~(\ref{eq:lagrangian}) by shifting the SM photon $A_\mu \to A_\mu - \epsilon A^\prime_\mu$ yields
\be
\label{eq:lagrangian-dark}
{\cal L} \to
-\frac{1}{4}F^\prime_{\mu\nu} F^{\prime\,\mu\nu} + \frac{m^2_{A^\prime}}{2} A^{\prime}_\mu A^{\prime\, \mu} + A^\prime_\mu (    g_D  J^\mu_D   +   \epsilon  e J^\mu_{\rm EM}  ) 
 ,~~~ J_{\rm EM }^\mu  = \sum_f Q_f \bar f \gamma^\mu f ,~~~~~~~~~~ (\rm Mass ~Basis) 
\ee
which induces an $\epsilon$ proportional coupling between $A^\prime$ and the EM current of SM particles $f$ with charges $Q_f$; the DM remains uncharged under the SM photon. 
  
 The phenomenology of the DM interaction depends on the DM/mediator mass hierarchy and on the details of the dark current $J^\mu_D$. If there 
 is only one dark sector state, the dark current  generically contains elastic interactions with the dark photon. However, if there are two (or more) dark sector states the dark photon can couple to the dark sector states off-diagonally, as we will illustrate shortly. This latter scenario can lead to distinct signatures, for which beam-dump experiments are especially suited.
 
 \begin{figure}[tp]
\center 
\includegraphics[width=16cm]{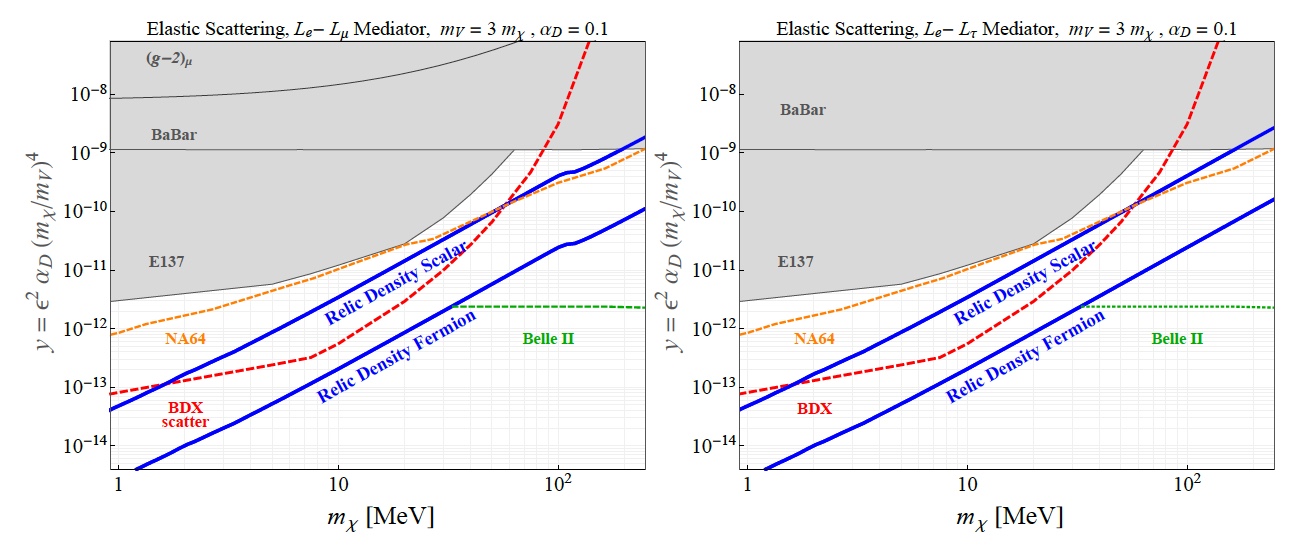} 

\caption{ Plot of BDX yield projections for elastic DM scattering $\chi e^- \to \chi e^-$ for  $10^{22}$ electrons on target 
mediated by {\it leptophilic} gauged $L_e - L_\mu$ (left) and $L_e - L_\tau$ (right) mediators. 
 For both panels, the blue curve represents the parameter space for which 
$\chi_1 \chi_2 \to \bar f f$ coannihilation yields the observed relic density and $f$ is a fermion from Eq.~(\ref{eq:Jemu}) or (\ref{eq:Jetau}).
The plots here based on the analysis in \cite{Izaguirre:2017bqb}, but computed specifically for this report with the mediators described in the text.  Also shown
are projections for NA64 and Belle II taken from the Cosmic Visions Report \cite{Battaglieri:2017CVR} (and rescaled for $\alpha_D = 0.1$)}
\label{fig:elastic}
\end{figure}

\subsubsection{Leptophilic  Mediators}
\label{sec:lepto}
Alternatively, instead of starting with a separate dark force $A^\prime$ and mixing with the SM photon (as above), we can couple both dark and visible matter 
to a new abelian 5th force, which gauges an existing combination of SM quantum numbers (e.g. baryon minus lepton number $B-L$). For  
 \be
 \label{eq:lagrangian-5th}
{\cal L} =
-\frac{1}{4}V_{\mu\nu} V^{\mu\nu}  + \frac{m^2_{V} }{2} V_\mu V^{\mu} +      V_\mu (        g_D J^\mu_D  +  g_{\rm SM} J^\mu_{\rm SM}  ) ,    
\ee
where $J_{\rm SM}$ is an anomaly free current of SM particles (not necessarily the electromagnetic current as in the dark photo scenario) and $V_{\mu\nu}  = \partial_\mu V_\nu  - \partial_\nu V_\mu $ is the fifth-force field-strength tensor. For comparison with the more familiar dark photon models, 
we define $\alpha_D \equiv g_D^2/4\pi$ as the dark fine structure constant and $\epsilon \equiv g_{\rm SM}/e$ as the SM coupling to $V$ normalized to the QED electron charge.

\begin{figure}[tp]
\center
\includegraphics[width=16.cm]{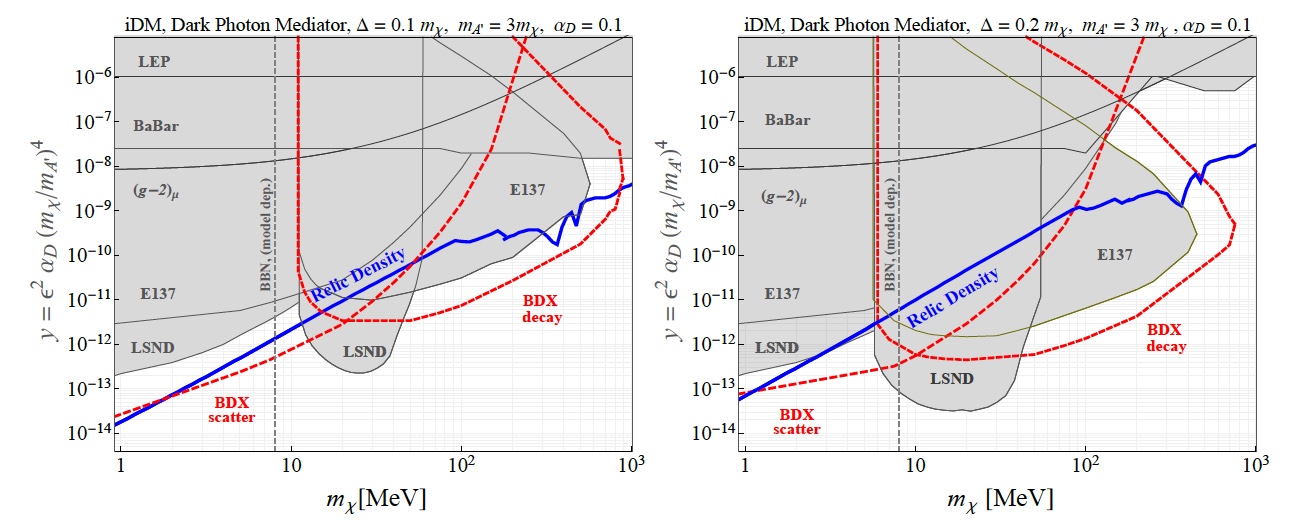}

\caption{ Plot of BDX yield projections for inelastic DM scattering $\chi_1 e^- \to \chi_2 e^-$ and decay $\chi_2 \to \chi_1 e^+e^-$ signatures mediated through a kinetically mixed dark photon mediator  (see Sec.~\ref{sec:dark-photon}) for $10^{22}$ electrons on target (red dashed curves). Two different mass splittings are shown: 10\% (left) and 20\% (right). Note that for the mass splittings considered here, direct detection scattering is forbidden
on kinematic grounds as the $\chi_1$ cannot upscatter to $\chi_2$ in non-relativistic collisions with SM targets.  
 In both panels, the blue dashed curve represents the parameter space for which 
$\chi_1 \chi_2 \to \bar f f, $ coannihilation yields the observed relic density, where $f$ is any charged SM particle; for details see \cite{Izaguirre:2017bqb}.      }
\label{fig:idm}
\end{figure}

Of particular interest are leptophilic mediators of the form $L_i - L_j$ which correspond to gauging a difference of two lepton flavors in the SM. Two representative
cases with appreciable electron couplings are gauged $U(1)_{L_e - L_\mu}$ for which the current  in Eq.~(\ref{eq:lagrangian-5th}) has the form
\be
\label{eq:Jemu}
J_{\rm SM}^\alpha =    \bar e \gamma^\alpha e  +  \bar \nu_e \gamma^\alpha \nu_e -
\bar \mu \gamma^\alpha \mu  -  \bar \nu_\mu \gamma^\alpha \nu_\mu   ~,~~~~~~~~{\rm Gauged ~} {L_e - L_\mu}
\ee
where, in addition to coupling to DM, the mediator $V$ also couples to all SM fermions with $e$ or $\mu$ flavor.
Similarly, we can write down the current for the gauged $L_e- L_\tau$  model where
\be
\label{eq:Jetau}
J_{\rm SM}^\alpha =    \bar e \gamma^\alpha e  +  \bar \nu_e \gamma^\alpha \nu_e -
\bar \tau \gamma^\alpha \tau  -  \bar \nu_\tau \gamma^\alpha \nu_\tau   ~,~~~~~~~~{\rm Gauged ~} {L_e - L_\tau}
\ee
where $V$ now couples to all SM fermions with $e$ or $\tau$ flavor. For the remainder of this document, we will emphasize these variations to highlight
 the characteristic features of leptonic forces; the other viable light mediators (e.g. $B-L$ or $B-3L_i$)  involve couplings to both quarks and leptons, so
 the parameter space is analogous to that of dark photons, which couple to the full EM current.

\subsubsection{Predictive Thermal Targets }
\label{sec:ann}
If the mediator MED = $A'$ or $V$ above is heavier than the DM, the thermal relic abundance is achieved via direct annihilation $\chi \bar \chi \to ff$, where $f$ are SM fermions. 
In the paradigm of a thermal origin for DM, DM would have acquired its current abundance through annihilation directly/indirectly into the SM. Here, we focus on the direct annihilation regime, in which $m_{\chi} < m_{\rm MED.}$, The annihilation rate to SM particles scales as 
\be
\hspace{1cm}({\rm direct ~annihilation},  ~  m_{\rm MED}  > m_\chi )~~~~ \sigma v_{       \chi \chi \to \bar ff }   \propto    y \equiv  \epsilon^2  \alpha_D  \left( \frac{  m_{\chi}}{   m_{\rm MED}  \!\!} \right)^4~~ ,~~~~~~~~~  ~~~~
\ee
and offers a  predictive target for discovery or falsifiability since the dark coupling $\alpha_{D}$  and mass ratio $m_{{\chi}}/m_{\rm MED}$ are 
at most ${\cal O}(1)$ in this $m_{\rm MED} > m_{\chi}$ regime, so there is a {\it minimum} SM-mediator coupling compatible with a thermal history.

\begin{figure}[tp]
\center
\includegraphics[width=16cm]{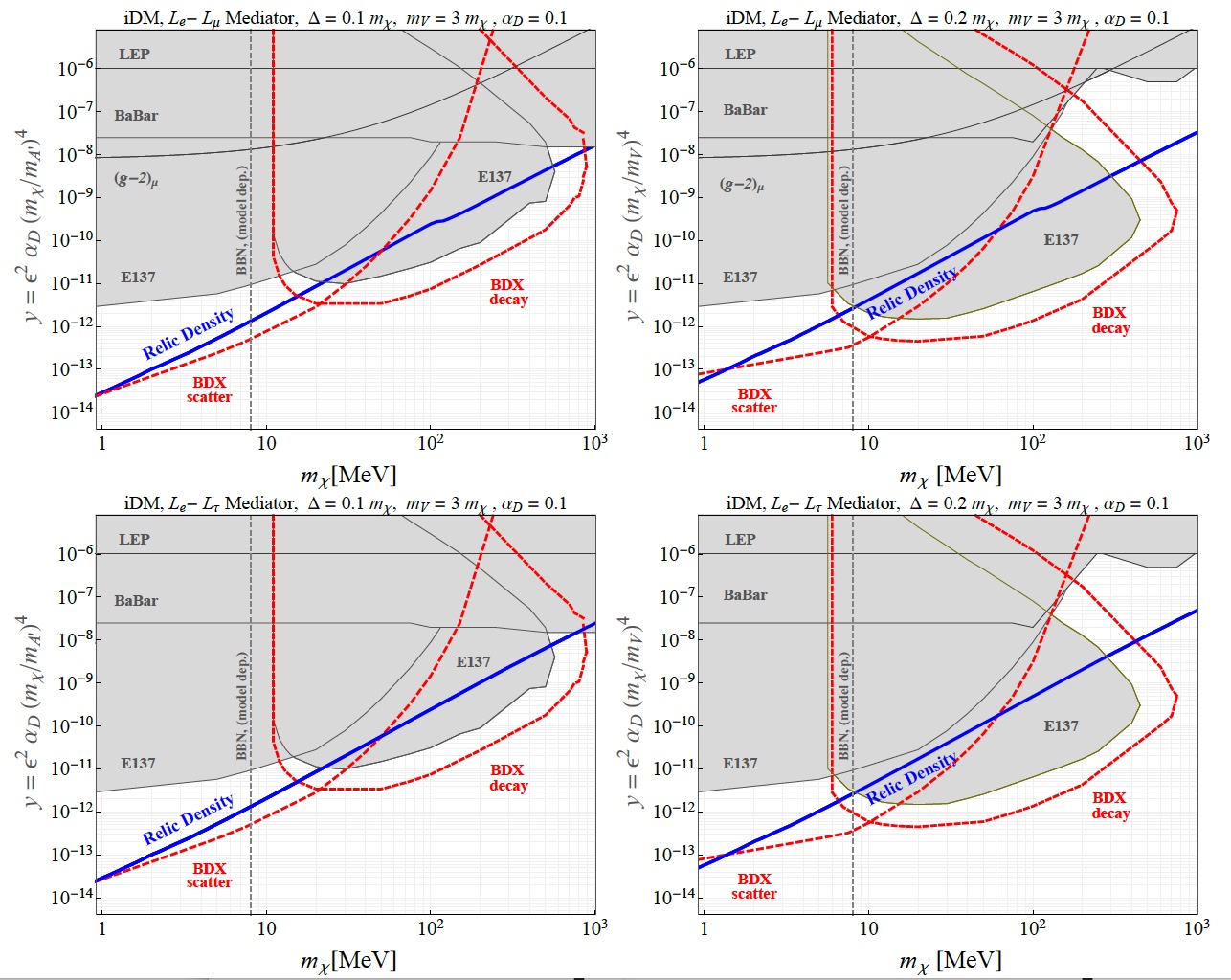} 

\caption{ Plot of BDX yield projections for inelastic DM scattering $\chi_1 e^- \to \chi_2 e^-$ and decay $\chi_2 \to \chi_1 e^+e^-$ signatures (red dashed) for  $10^{22}$ electrons on target 
mediated by {\it leptophilic} gauged $L_e - L_\mu$ (top row) and $L_e - L_\tau$ (bottom row) mediators. 
 Note that for the mass splittings considered here, direct detection scattering is forbidden
on kinematic grounds as the $\chi_1$ cannot upscatter to $\chi_2$ in non-relativistic collisoins with SM targets.  
 Two different mass splittings are shown: 10\% (left column) and 20\% (right column). In all four plots, the blue curve represents the parameter space for which 
$\chi_1 \chi_2 \to \bar f f$ coannihilation yields the observed relic density and $f$ is a fermion from Eq.~(\ref{eq:Jemu}) or (\ref{eq:Jetau}).
The plots here are based on the analysis in \cite{Izaguirre:2017bqb}, but computed specifically for this report with the mediators described in the text. }
\label{fig:idm-lepto}
\end{figure}

\subsubsection{Dark Matter Candidates}
\paragraph{Scalar Dark Matter (Elastic Scattering)}
If DM interacts elastically with the SM, we can write down benchmark models
for which the dark current couples to the mediators in Eqs.~(\ref{eq:lagrangian-dark})  or Eq.~(\ref{eq:lagrangian-5th}) as 
\be
J_{\rm DM}^\mu =
 i  (\chi^* \partial_\mu \chi - \chi \partial_\mu \chi^* )~~,~~ (\rm Scalar ~DM) 
\ee
where the thermal target is shown in Fig.~\ref{fig:elastic} alongside
various bounds and BDX reach projections for the leptophilic mediators introduced in Sec.~\ref{sec:lepto}. Note that the scalar target requires larger couplings than the pseudo-Dirac fermion 
target because $\sigma v \propto v^2$ is $p$-wave, so there is a mild velocity suppression of the annihilation
cross section during freeze-out, which must be compensated with slightly larger couplings. This model is safe
from CMB bounds on DM annihilation during recombination because the $p$-wave cross section is 
significantly smaller at later times.

\paragraph{Pseudo-Dirac Dark Matter  (Quasi Elastic Scattering)}
If four-component fermion $\chi$ has a sizable Dirac mass and a small Majorana mass through $U(1)$ symmetry breaking, which
gives the mediator its mass, then the fermion mass terms become
\be
{\cal L}_{\rm mass} = m_D \bar \chi \chi + m_M \bar \chi^c \chi  ~~~~~{(\rm Interaction ~Basis) }~~\to ~~~
~ m_1 \bar \chi_1 \chi_1 + m_2 \bar \chi_2 \chi_2 ~~~~~~ {(\rm Mass ~Basis) }
\ee
where $m_{D,M}$ are respectively the Dirac and Majorana masses and $^c$ denotes charge conjugation. In the mass basis, the eigenstates of this system $\chi_{1,2}$ are split in 
mass by a small amount $\Delta \equiv m_2 - m_1$ and the leading interaction with the mediator is off-diagonal 
\be
\label{eq:idm-current}
 J^\mu_{\rm DM}  = \bar \chi \gamma^\mu \chi      ~~~~ { (\rm Interaction~ Basis)}   ~~  \to ~~~ \bar \chi_1 \gamma^\mu \chi_2 + h.c.~~~~ ~~(\rm Mass~Basis) ~,~
\ee
so the dark matter candidate $\chi_1$ (which we refer to as simply $\chi$) must inelastically upscatter into the heavier state $\chi_2$ in order
to interact with SM particles. This class of models is safe from CMB bounds because the annihilation stops once the heavier, unstable $\chi_2$ decay in the 
early universe. In Fig.~\ref{fig:elastic} we show the Pseudo-Dirac thermal target alongside the scalar target with 
corresponding bounds and projections for leptophilic mediators.

\paragraph{Inelastic Dark Matter -- iDM (Scattering and Decays)}
In the limit where the mass splitting  $\Delta \equiv m_2 - m_1$ of a Pseudo-Dirac particle becomes appreciable with respect to the Dirac mass ($\sim$ 10s of $\%$),
the same current in Eq.~(\ref{eq:idm-current}) allows for the possibility of $\chi_2 \to \chi_1 \bar f f$ decays, as long as the mass splitting
between the ground and excited state satisfies $\Delta < m_{\rm MED}$ where MED is either the dark photon $A^\prime$ or 5th
force gauge boson. Unlike in the Pseudo-Dirac limit, now thermal freeze-out can be significantly affected by the mass difference between
coannihilation partners $\chi_{1,2}$, such that larger splittings also require {\it larger couplings} to achieve the observed DM density. 

The BDX signature from $\chi_2$ decays is qualitatively different as it scales with the detector volume and arises from $\chi_2$ passing through the detector
after being produced in the beam dump from $A^\prime$ decays. In Fig.~\ref{fig:idm} we show existing bounds and BDX projections for 
iDM scenarios with a dark photon mediator and various choices for $\Delta$. Similarly, in Fig.~\ref{fig:idm-lepto} we show related benchmarks
for leptophilic $L_e -L_\mu$ and    $L_e -L_\tau$ mediated scenarios introduced in Sec.\ref{sec:lepto}.

\subsection{BDX Complementarity \label{sec:complementarity}}
There are several other experimental concepts that aim to cover the parameter space for light thermal dark matter. These 
concepts were comprehensively outlined in 2017 DOE Cosmic Visions Report (CVR) \cite{Battaglieri:2017CVR}. These  proposals
fit into two main categories:
\begin{itemize}
\item Section IV of the CVR presents
new ideas for proposed direct detection experiments involving novel detector targets with varying degrees
of technological maturity(e.g. semiconductors, superconductors, graphene, etc.). For elastic scattering DM
models with electron couplings, these efforts aim to probe some of the parameter space that BDX is projected
to cover, but are insensitive to models with inelastic interactions (e.g. Pseudo-Dirac DM and iDM described above), which
whose potential scattering processes are kinematically forbidden. Similarly, any model (e.g. Majorana DM) whose non-relativistic 
scattering cross section is velocity dependent, and therefore prohibitively suppressed at direct detection experiments. 
\item Section VI  of the CVR presents new accelerator based ideas for studying light dark matter, independently of its local abundance in
the Galactic halo. These ideas involve both electron beams (BDX, LDMX, NA64, Belle II) and proton beams (SBND, COHERENT, MiniBooNE, SHiP).
Unlike direct detection experiments, these techniques are independent of astrophysical uncertainties and 
all rely on relativistic production and/or detection, thereby minimizing the relative differences between the lorentz structures
of various SM/DM interactions -- e.g. the direct detection sensitivity depends significantly on the spin or velocity dependence of 
interaction rates, wheras accelerator based experiments (where $v \sim c$) only differ by order-one amounts. 
Furthermore, unlike dark force searches, which look for resonances corresponding to visibly decaying particles, these efforts primarily aim to 
study invisibly (or partly invisibly) decaying particles.
\end{itemize}

Unlike other efforts, BDX is the only proposed experiment that features both DM production and detection utilizing  
only its coupling to electrons. It can therefore test viable models which do not require  any couplings to baryons (which SBND, COHERENT, MiniBooNE, and SHiP all
require) and it can directly observe the DM scatter (or iDM decay) in the downstream detector (which LDMX, NA64, and Belle II cannot). Furthermore, BDX
is unique even among other electron based approaches in its ability to observe the particles produced in the fixed-target, 
whereas other electron-beam experiments rely on missing energy signatures,
which indirectly infer the production of signal events.  

To compare different electron beam strategies, in Fig. \ref{fig:elastic}
we show parameter space for leptophilic, elastically coupled DM in the $y$ vs $m_\chi$ parameter space alongside thermal targets for representative models. 
Also shown are projections for BDX with $10^{22}$ electrons on target (EOT), NA64 with $10^{11}$ EOT, and Belle II with 50 ab$^{-1}$ of luminosity taken from the 
Cosmic Visions Report~\cite{Battaglieri:2017CVR}. Since NA64 is an existing experiment at CERN, this projection is reliable and realistic; it only assumes additional data 
taking with an established experimental method. Although previous $B$-factory searches at BaBar have demonstrated that 
such $e^+e^- \to $ mono$-\gamma$ searches are powerful probes of light dark matter,  the
 Belle II projection from the CVR is not verifiably robust depends on self-reported analyses
submitted to the relevant CVR working groups; there is no peer-reviewed publication that has validated this projection. By contrast, the BDX projections
shown here are based on rigorous background estimates, realistic data-driven detector efficiencies, and achievable CEBAF luminosities. Furthermore, 
the beam dump production and scattering technique has been demonstrated as a powerful probe of light DM by 
 reinterpretations of old E137 data~\cite{Batell:2014mga}, which relied on the same basic (but suboptimal) experimental setup. 

The CVR also discusses the proposed LDMX experiment, which aims to study fixed target missing momentum production of invisibly decaying dark forces. 
The LDMX projections shown in Fig. 20 of Sec VI assume a 3 event yield (not a sensitivity projection) for a futuristic Phase 2 run with $10^{16}$ EOT, whose
feasibility has not yet been demonstrated; this curve should not be regarded as direct competition for BDX until it can be shown that Phase
2 can reach a negligible background level. The LDMX collaboration is currently preparing a document to present Phase 1 projections for a
 4 GeV beam with $4 \times 10^{14}$ EOT, which may cover some portion of the BDX parameter space. However, this approach has a 
 cost estimate in the \$ 10M range and -- unlike BDX, NA64, or Belle II -- it relies on a currently untested technology. Thus,  it is not ultimately clear  whether it will have an impact on a relevant timescale, especially with competition from currently running experiments (NA64) and proposed experiments based on techniques (BDX, Belle II).
 
\subsection{Summary of theoretical update}

In summary, BDX is uniquely suited to make timely and significant progress towards testing predictive models of sub-GeV DM. It is the ultimate electron beam-dump experiment, whose sensitivity will only be limited by irreducible beam-related neutrino floor. Unlike any other experiment currently being proposed, BDX can  both produce and detect
light DM exclusively through its interaction with electrons. For this reason, the experiment is qualitatively distinct even from other electron-beam missing energy 
strategies in its ability to directly study the DM itself through its interactions with the downstream detector. For nominal luminosities, it has been demonstrated
that BDX can reach important theoretical milestones that will test predictive models of light dark matter coupled to dark photons and other light new 
leptophilic forces.

%% file: BDX-PAC46-mutest.tex
\section{Beam-on background assessment}
\label{sec:mutest}
\begin{figure}[tph] 
\center  
\includegraphics[width=10.5cm]{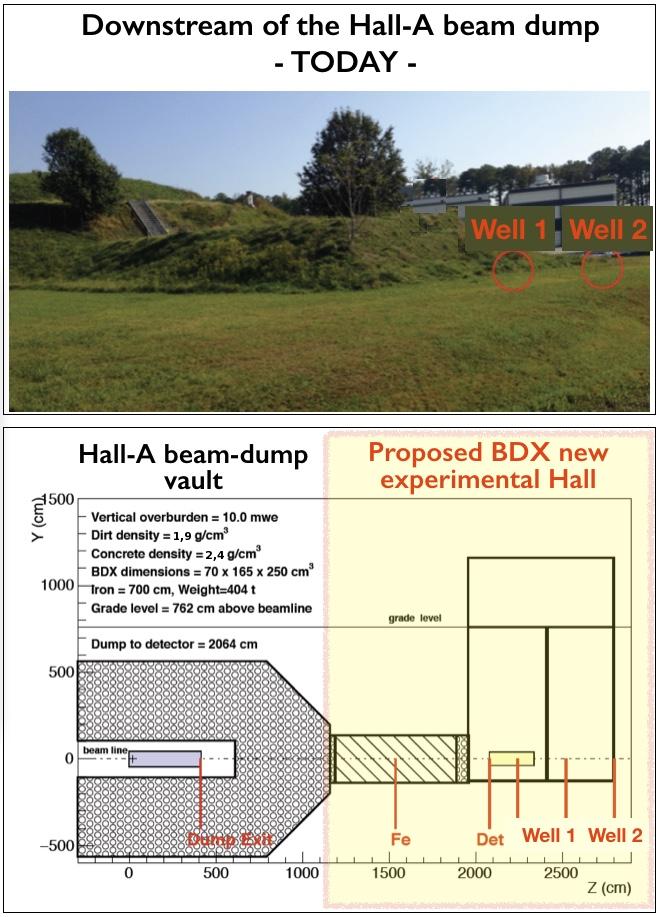}     
\caption{The area downstream of the Hall-A beam-dump and the studied test locations. From left to right: beam-dump  exit, iron shielding, BDX detector  front face, Well-1 and Well-2.
\label{fig:pipe-location}}
\end{figure} 
A complete  beam-on background assessment in the BDX detector, will only be possible when the new underground facility (including the additional iron shielding between the dump and the detector) will be  built. In the current configuration, the radiation produced by the interaction of the 10.6 GeV electron  beam on Hall-A beam dump is only partially shielded by the dump vault concrete and $\sim$20 m of dirt. As proposed to and supported by PAC45, in spring 2018 we measured the muon flux  in test pipes located behind Hall-A at beam height.
The results of this test were used to validate  Montecarlo simulations and confirm that no other sources of unpredicted background are present (e.g. T$_N>$10 MeV neutrons). In this Section  we report details of the measurement.

\subsection{Test set-up}
\subsubsection{Location of measurement pipes }
\begin{figure}[tph] 
\center  
\includegraphics[width=10.cm]{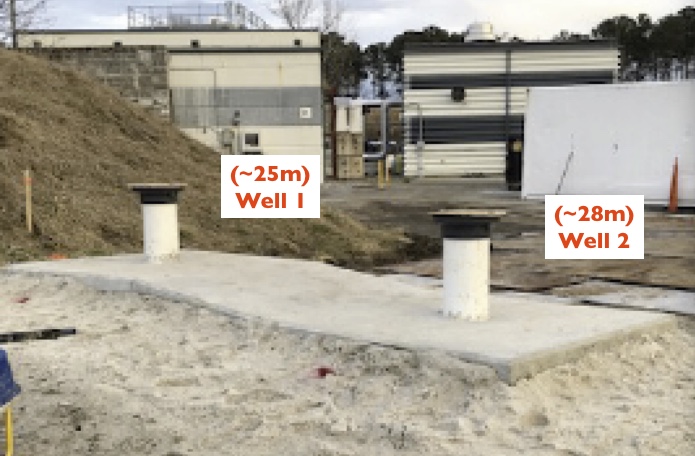}     
\caption{The two pipes installed.
\label{fig:wells}}
\end{figure} 

The area downstream of Hall-A beam-dump is shown in Fig.~\ref{fig:pipe-location}  indicating test measurement locations relative to the 
new  underground facility proposed in PR12-16-001~\cite{bdx-proposal}. Two wells have been dug in the positions marked as 'Well-1' and 'Well-2'.  Two pipes, 10" in diameter and 9m in length, were installed approximately 25 m and 28 m downstream of the dump.  Figure~\ref{fig:wells} shows the two pipes installed over  the concrete platform.
\begin{table}[h!]
\centering
\begin{tabular}{|c|c|c|c|}
\hline
Component & X (cm) & Y (cm) & Z (cm)  \\
\hline
Hall A center & 0 & $0$ & 0\\
\hline
Entrance of tunnel & 0 & $0$ & 2652\\
\hline
Beam diffuser & 0 & $0$ & 3276\\
\hline
Beginning of dump & 0 & $0$ & 4996\\
\hline
End of dump & 0 & $0$ & 5316\\
\hline
End of tunnel & 0 & $0$ & 5571\\
\hline
Well-1 at elev 13.5 ft (beam height)& -1.3 & $ -3.5 $ & 7569\\
\hline
Well-2 at elev 13.5 ft (beam height) & 3.8& $ -3.0 $ & 7874\\
\hline
Well-1 at elev 36.6 ft (top of Al plate)& 0.6 & 791.9 &  7572\\
\hline
Well-2 at elev 36.6 ft (top of Al plate)& 1.6 & 777.2 &  7879\\
\hline
\hline
\end{tabular}
\caption{\label{tab:pipe-pos} Position of the different components of Hall-A beam line and BDX pipes. Positive X corresponds to North direction;  positive Y corresponds  up direction; positive Z is along the beam direction. }
\end{table}
The  precise localization of the two wells (with respect to the Hall-A beam dump) was established by two independent surveys performed by JLab Facility and Survey groups~\cite{survey-priv-com,survey-priv-com1}.  Results  are reported in Tab.~\ref{tab:pipe-pos}. The systematic error associated to the procedure has been estimated to  be $\Delta Pos = \pm$ 5 cm.
A tent was mounted around the two pipes to protect the DAQ system and facilitate field operations.
During the test, the BDX-Hodo detector (see below) was  lowered in the pipe and the muon flux sampled at different heights with respect to nominal beam height. The muon flux profiles in Y (vertical direction), measured in the two  different locations in Z (distance from the dump),  allowed us to compare the absolute and relative MC predictions. 

\subsubsection{The BDX-Hodo detector \label{sec:BDX-hodo}}
The detector used to measure the beam-on-related muon radiation and the background in the proximity of the new BDX underground facility was made by  a BDX ECal CsI(Tl) crystal, identical to the ones proposed for the full experiment, sandwiched between a set of segmented plastic scintillators.
\begin{figure}[ht!] 
\center 
\includegraphics[width=12.cm]{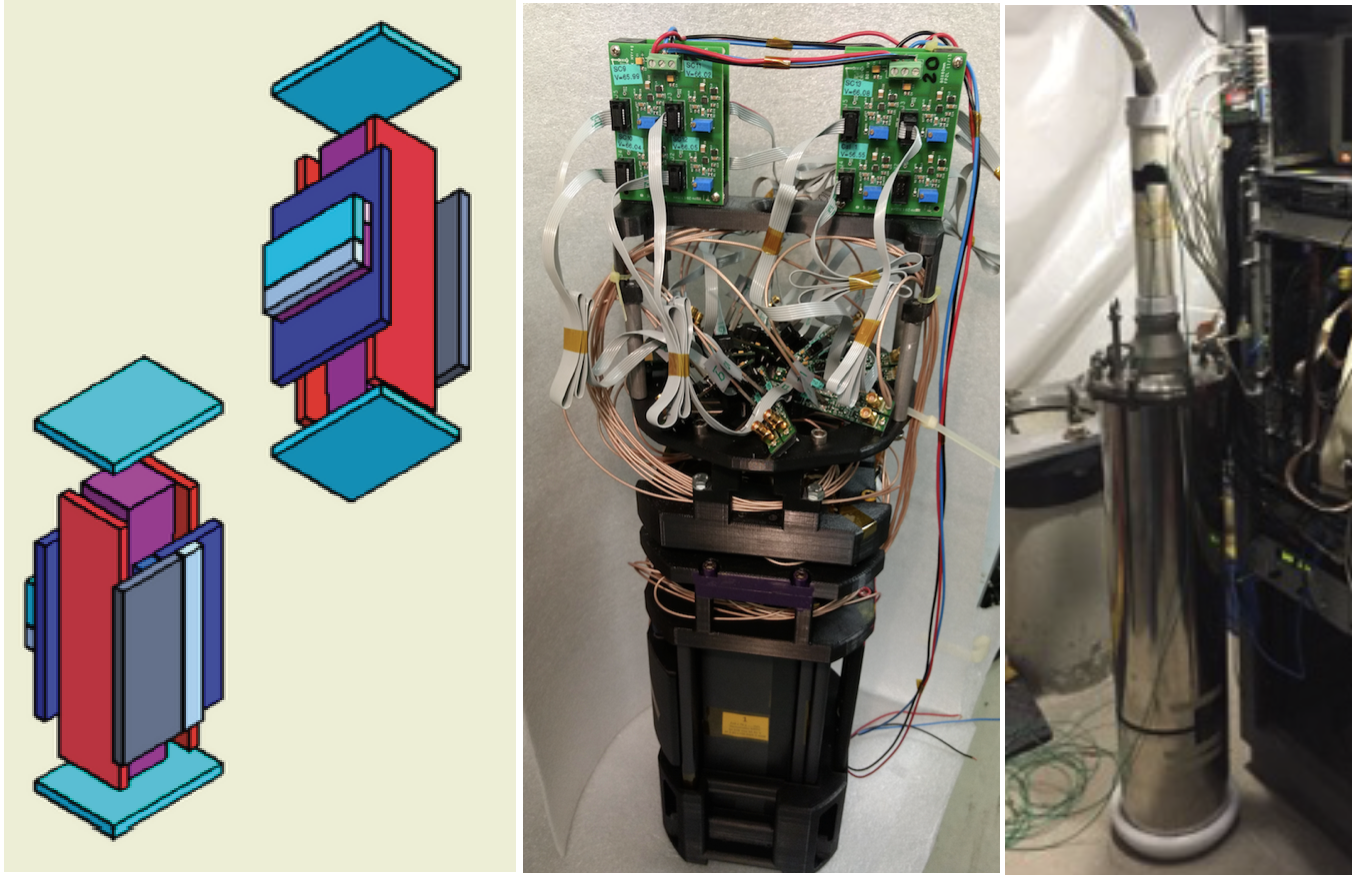}  
\caption{The CAD representation of the BDX-Hodo detector (left) and some pictures of the detector assembled.
}
\label{fig:det-cad}
\end{figure}
The detector was assembled with technologies proposed for use in the final experiment so it has similar sensitivities to background.
The requirement of a hit in both front and back paddles defines a 3x3 matrix of 2.5x2.5 cm$^2$ pixels providing cm-scale muon XY position resolution. 
Four more  paddles covering  the left/right sides and the top/bottom of the crystal were used to (partially) veto cosmic rays and other radiation not coming from the beam direction.
 The scintillator paddles were  made with clear plastic, each  read out via a WLS fiber coupled to a 3x3 mm$^2$ Hamamatsu S12572-100 SiPM sharing the same technology used in the BDX Inner Veto detector (described in details in Sec. 3.2.2 of  PR12-16-001).
The detector was contained  in a 20-cm diameter stainless-steel cylindrical vessel, covered on top and on the bottom by steel lids. The whole assembly is   water-tight to prevent any water from leaking inside the vessel. A PVC  extension on the top cover was  used  to run  cables (signal and power)  from the detector to the ground-level and to orient the detector in the direction of the beam. 
Data acquisition was triggered by the ``OR'' of different logical conditions from scintillators and crystal. 
The  DAQ system (VME crate + pc) was hosted in the tent. It was connected to JLab Accelerator network via an optical link to facilitate the raw and analyzed data transfer to the JLab CC Silo. An on-line monitoring program allowed us to have prompt feedback on rates recorded by  scintillators and crystal scalers.
The DAQ live-time was kept $LT>85\%$ by pre-scaling high rate triggers. Live-time information, as well as the trigger bits prescale factors,  were merged into the data stream and used in the off-line analysis to properly calculate event rates.
\subsubsection{Run conditions}
Data were collected from February 22nd to May 2nd 2018 with Hall-A beam parameters (for most of the time) reported below:
\begin{itemize} 
\item{$I_{Beam}= 22\mu$A,}
\item{$E_{Beam}= 10.6$ GeV.}
\end{itemize} 
Some special runs were taken at different beam currents (2.2$\mu$A, 5$\mu$A, and 10$\mu$A) and, for   a week, data  were recorded  at $E_{Beam}= 4.3$ GeV.
We collected about 100 (short) runs for a total of about 10 TByte of data (raw + cooked) processed locally and transfered to the JLab CC silo and $\backslash volatile$ storage area for further analysis.
The Hall-A  beam EPICS scaler information was merged into the data stream for off-line analysis. Time periods corresponding to beam trips were cut out and not used in the analysis. Other parameters monitored during the data period,  such as ambient temperature and pressure, were recorded  and made available off-line.

\subsection{Results}
\subsection{Muon selection}
\begin{figure}[ht!] 
\center 
\includegraphics[width=12.cm]{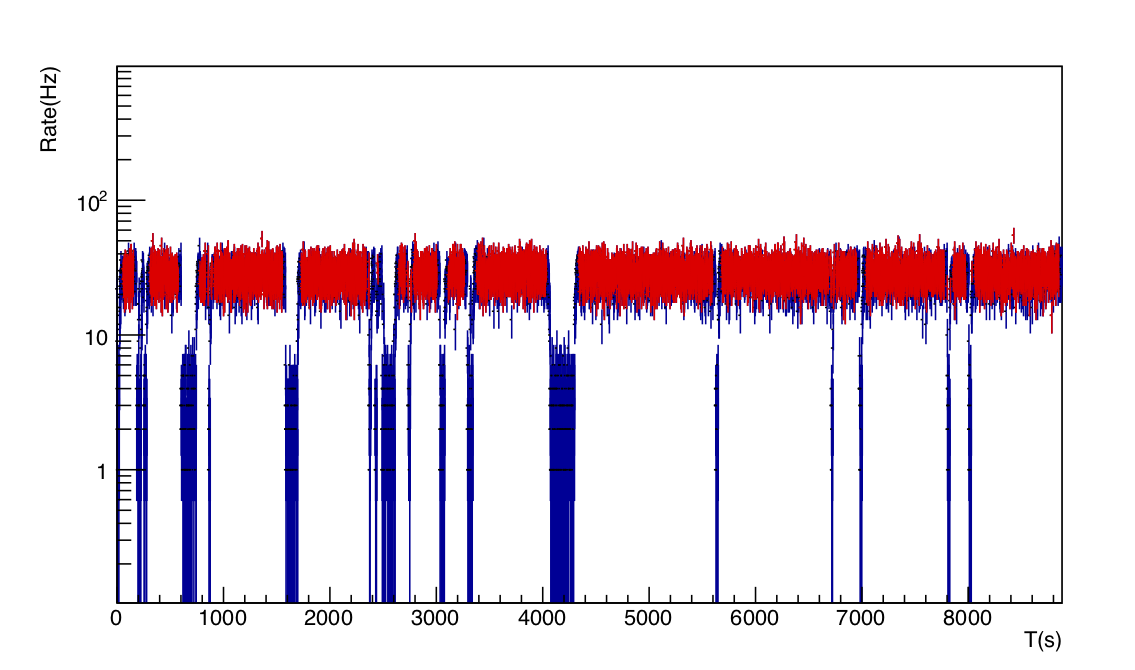}  
\caption{Example of rate measured in the crystal as a function of time. Rate going to zero corresponds to beam-trip events. The full data set is shown in blue. Only data acquired during stable beam-current conditions (red points) have been included in the data analysis.  
}
\label{fig:CrsRATE_vs_time}
\end{figure}

In the the off-line analysis, data have been 
cleaned,
by excluding the beam-trip events, and calibrated. 
As an example, the measured rate in the crystal as a function of time  is reported in Fig.~\ref{fig:CrsRATE_vs_time}. 
As far as the energy calibration is concerned, we used the single photo-electron peak  recorded in special pulser runs,  to calibrate  plastic scintillator SiPM's and cosmic muons to establish the absolute energy scale for the CsI(Tl) crystal.  

\begin{figure}[ht!] 
\center 
\includegraphics[width=17.cm]{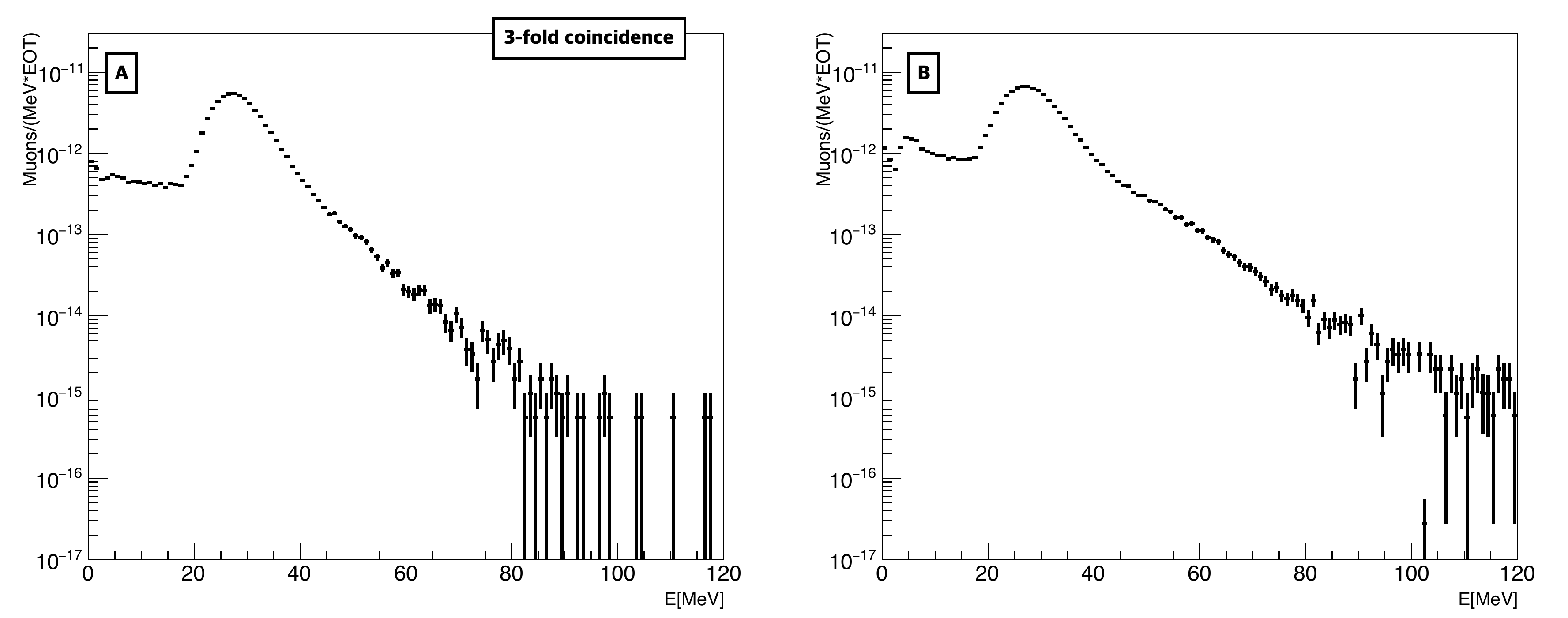}  
\caption{Measured event rate as a function of deposited energy in the crystal with (A) and without (B)the 3-fold coincidence.  
}
\label{fig:Crs_spectra}
\end{figure}

\begin{figure}[ht!] 
\center 
\includegraphics[width=12.cm]{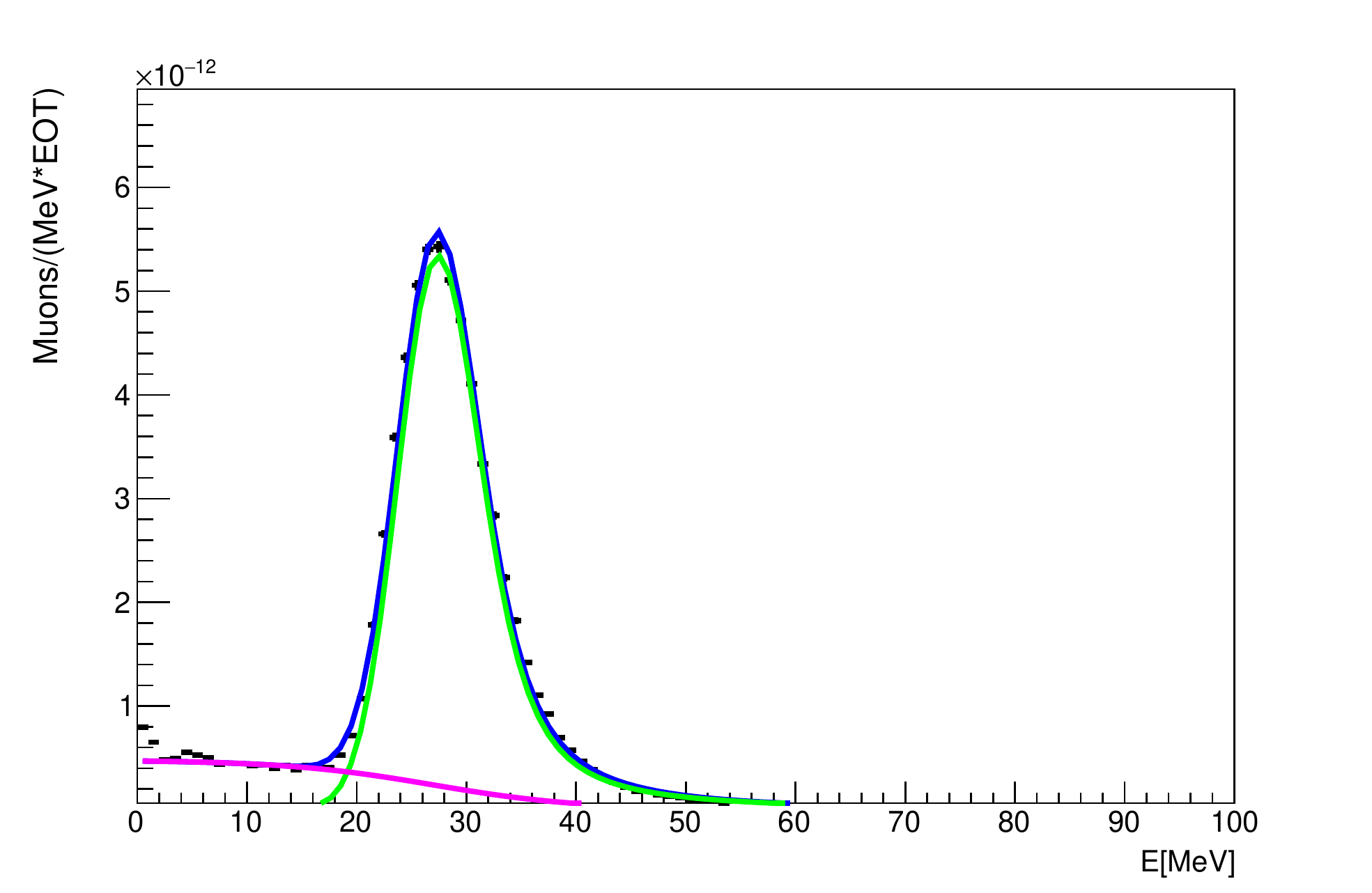}  
\caption{Example of a Maximum Likelihood fit (blue) to the CsI energy spectrum to extract the rate of beam-related muons crossing the crystal. The fit has been performed modeling the background (pink) with a Fermi function. The muons/EOT rate has been obtained by integrating the signal (green) in the full energy range.
}
\label{fig:FCrsB_fit}
\end{figure}

Muons have been selected by requiring a 3-fold coincidence between crystal, front and back paddles. An example of the muon rate normalized to the beam current as a function of the deposited energy in the crystal is shown in Fig.~\ref{fig:Crs_spectra}. The crystal energy spectrum without coincidence conditions is also reported in the same figure.  
The crystal energy spectrum was fitted to a Landau function convoluted with a Gaussian (to take into account the detector resolution), modeling the low energy background with a Fermi function, as shown in Fig.~\ref{fig:FCrsB_fit}.   
\subsection{Muon rates}
The muon rates were obtained by integrating the Landau function from 0 MeV to 120 MeV. The systematic error in background parametrization represents the main source of uncertainty in the extracted rates. It is included in error bars in all  
 following plots.

\begin{figure}[tp] 
\center 
\includegraphics[width=8.cm]{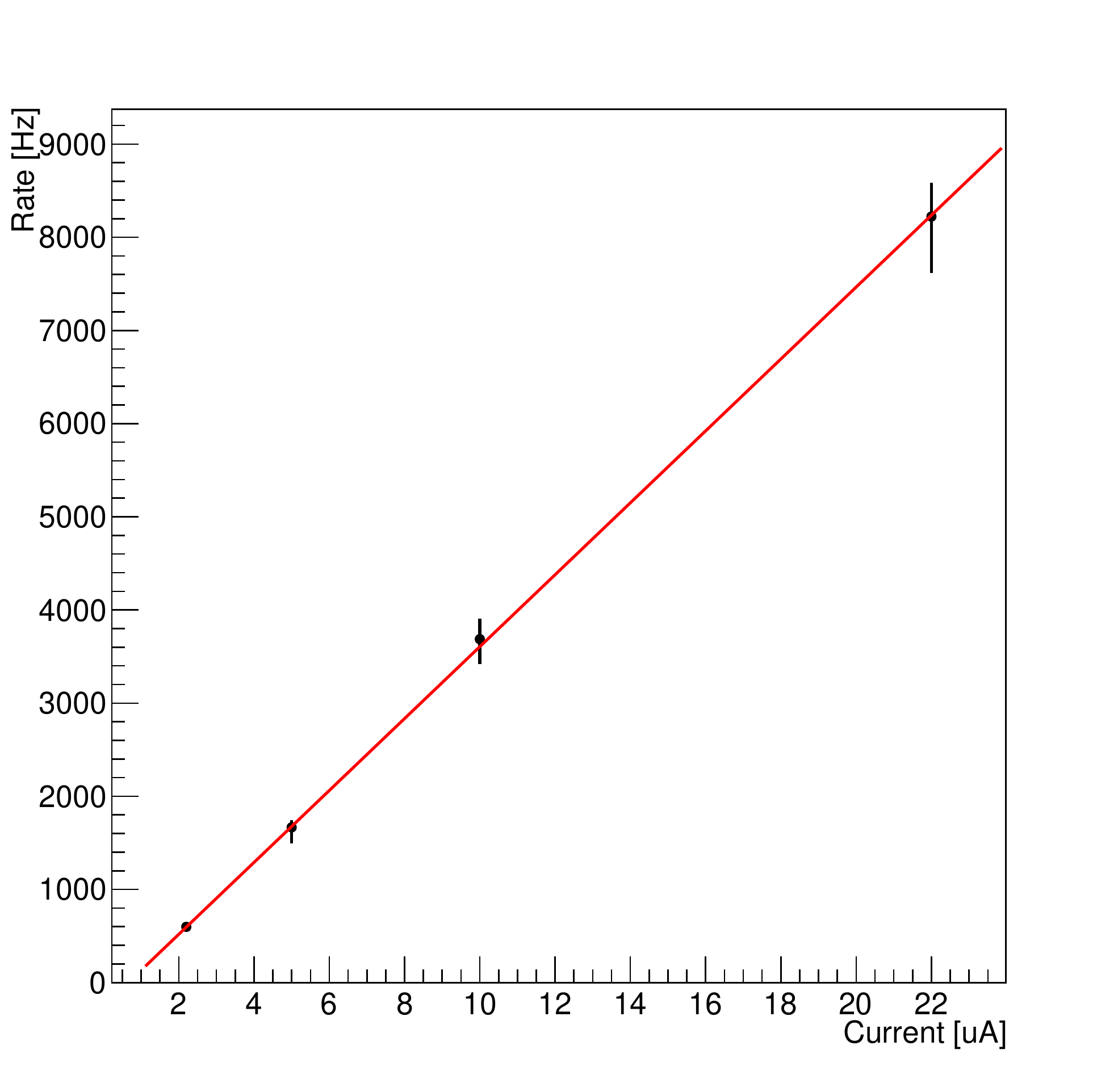}  
\caption{Muon rate measured inside the well-1 at the beam line height as a function of the beam current. The red line is the linear best-fit of the data.
}
\label{fig:rate_vs_current}
\end{figure}

As a check,  we measured the correlation between the beam current and the muon rate for a fixed detector position. Figure~\ref{fig:rate_vs_current} shows the rate measured with the detector inside Well-1 at the beam-line height for 4 different beam current values: 2.2 uA, 5 uA, 11 uA, and 22 uA. 
As expected, the muon rate dependence on current is linear.

\begin{figure}[tp] 
\center 
\includegraphics[width=7.8cm]{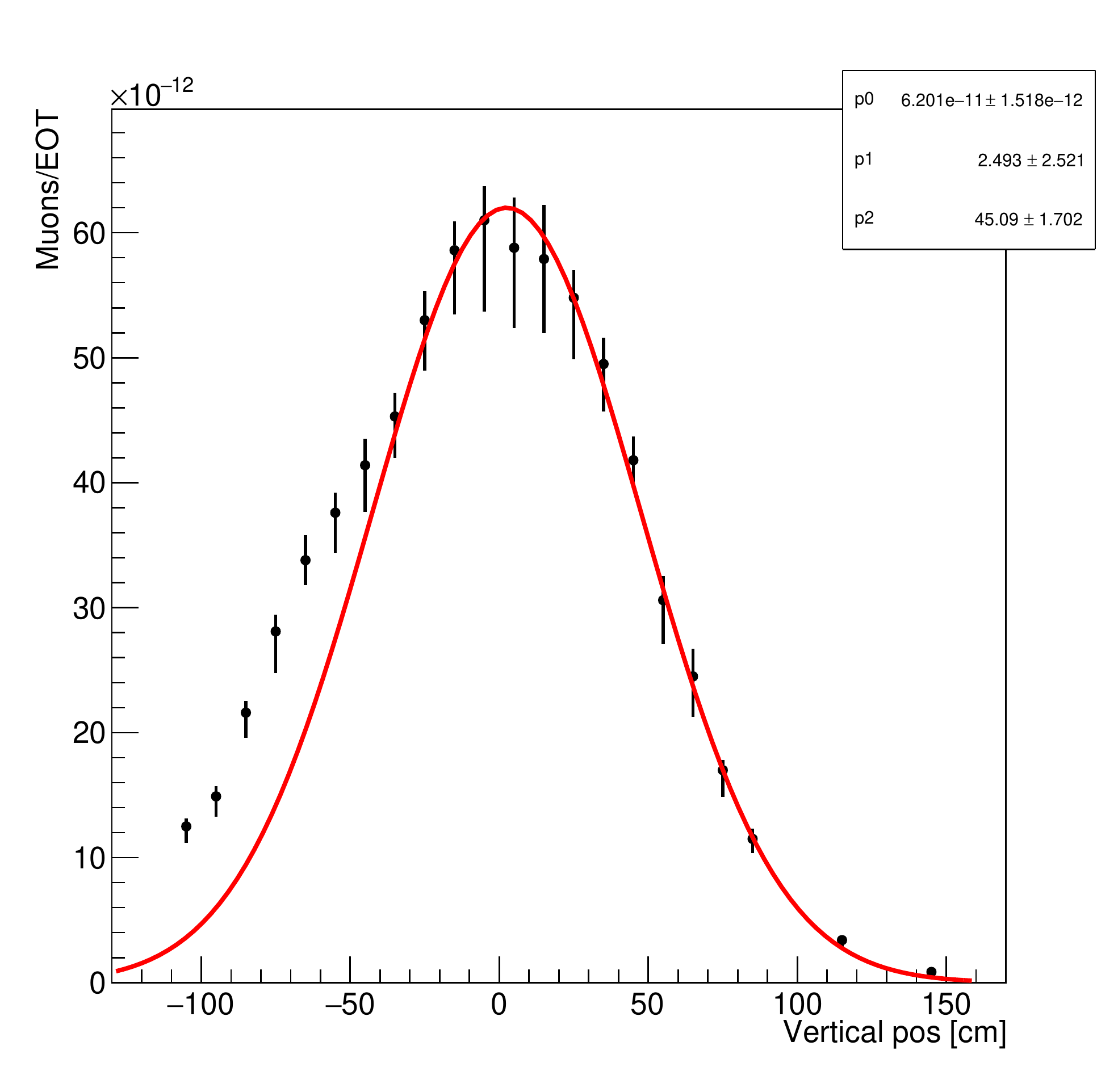}  
\includegraphics[width=7.8cm]{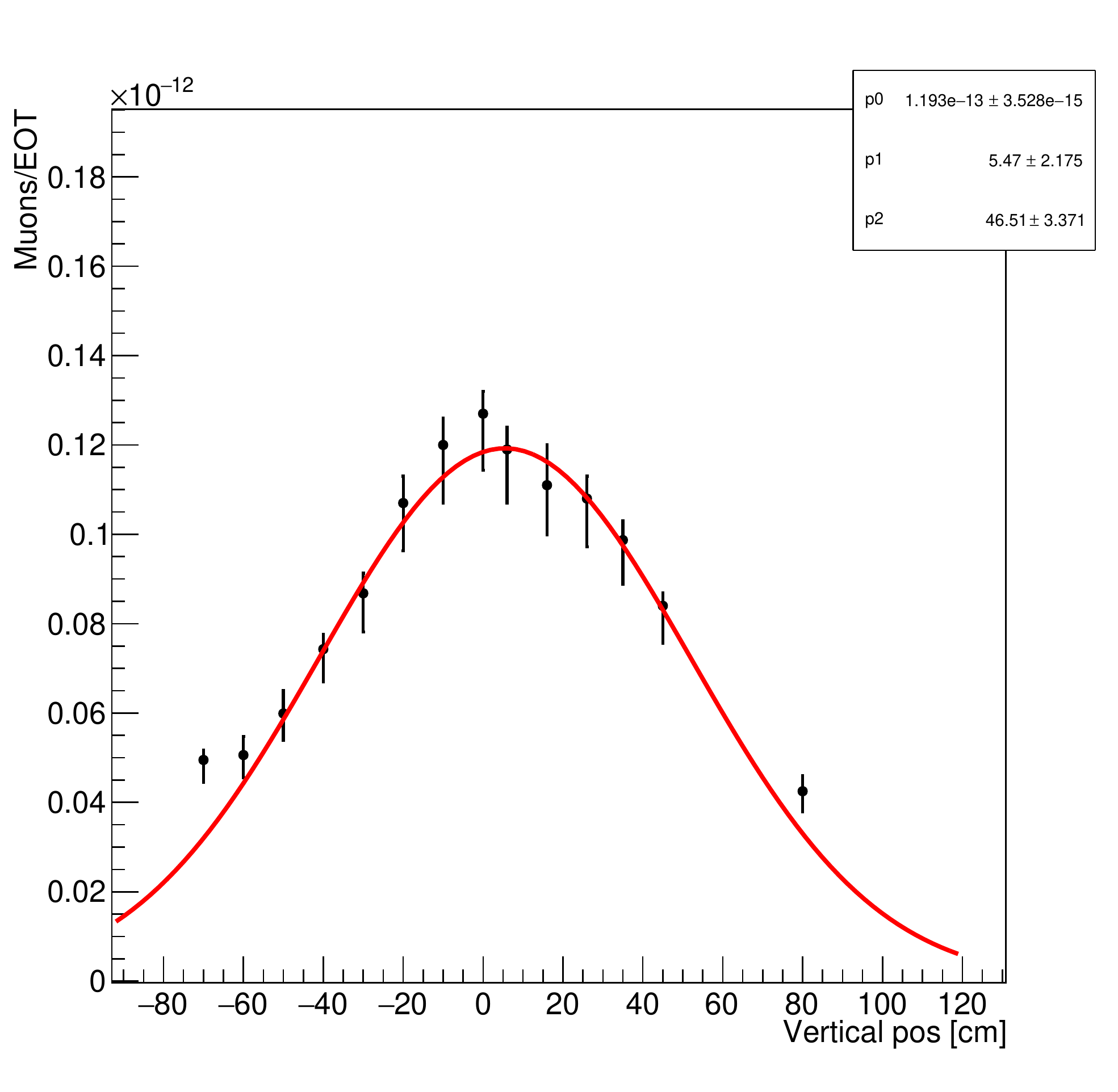}  
\caption{Muon rate as a function of the detector vertical position measured in Well-1 (left) and Well-2 (right). The red line is the Gaussian best-fit of the data. 
}
\label{fig:1-2wel_rate}
\end{figure}

The muon flux sampled at different heights with respect to the beam line for  Well-1 and Well-2 are shown in Fig.~\ref{fig:1-2wel_rate}. Positive (negative) position values refer to detector positions above (below) the beam-line. 
In the figures the distributions have been centered to zero shifting by -10cm (Well-1) and -40cm (Well-2) the positions where the maximum rate was measured. 
Whereas the slight shift observed in Well-1 is compatible with the detector position systematic uncertainty ($\Delta Pos = \pm$ 5 cm), there is not a clear explanation for the size of the  shift observed in Well-2. 
The maximum rates measured in Well-1 and Well-2 were found to be 8.4 KHz and  17 Hz, respectively.
Distributions of rate as a function of the distance from the nominal  beam height in the two wells show a nearly symmetric shape around the maximum. They   were fitted to  Gaussian functions finding a similar widths  of $\sigma$ = (45  $\pm$ 2) cm and $\sigma$ = (46  $\pm$ 3) cm for Well-1 and Well-2 respectively. For both wells, data points in the range  -50 cm to -100 cm show a deviation from the gaussian line with an excess of counts. This may be due to variations (a reduction) in the soil density traversed by incoming muons below beam height.

\subsection{Comparison to simulations}
We used GEANT4 and FLUKA to simulate the interaction of 10.6 GeV electron beam with Hall-A beam dump. Details of the  procedure, comparison between the two programs and expected  rates in the well's location proximity, were reported in~\cite{bdx-update-PAC45}. From this study, we concluded that over a certain energy  threshold, around  the MIP's  peak ($\sim$25 MeV), only muons would be detected by BDX-Hodo, with rates in the range 10 Hz - 10 kHz depending on the distance from the dump and the beam-height \footnote{Neutrons were expected to populate the low energy part of the spectrum with negligible  contribution for energies greater than  20 MeV.}.
\subsubsection{The experimental conditions}
The precise position of the two pipes, as determined from the JLab survey, as well as details of the Hall-A run conditions (current, energy and beam diffuser)  were included in the simulation.

An important parameter is the composition and density of materials crossed by muons  while traveling from the dump to the detector (mainly dump-vault concrete and soil). During the excavation, two soil samples were taken near the pipe locations resulting in    $\rho_{dirt}=$1.93 g/cm$^3$ and  1.95 g/cm$^3$, respectively. We could not sample the concrete and therefore we assumed $\rho_{concrete}$ in the range (2.2 - 2.4) g/cm$^3$, as suggested by JLab Facility Group.
\begin{figure}[tph] 
\center 
\includegraphics[width=8.2cm]{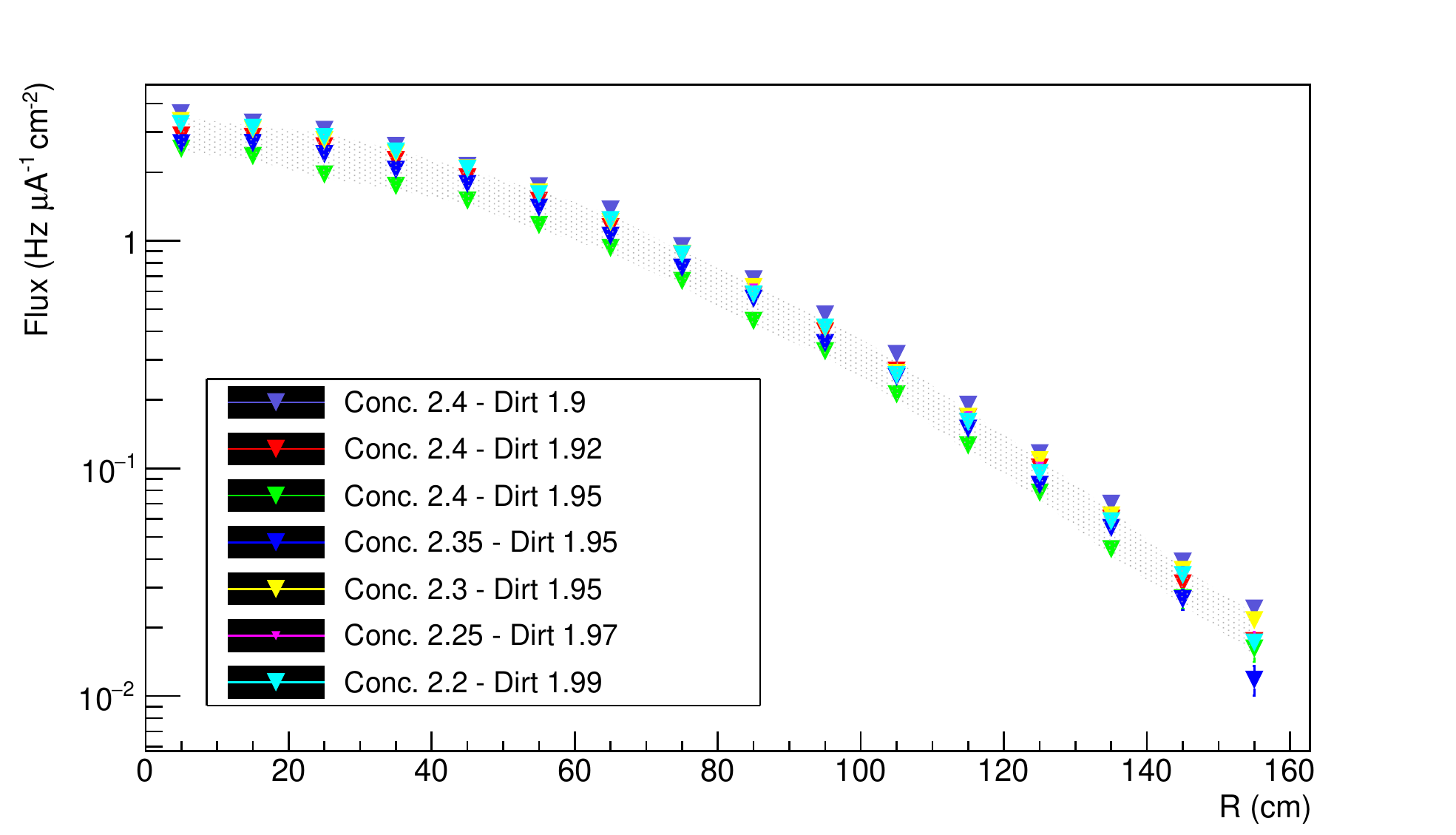}  
\includegraphics[width=8.2cm]{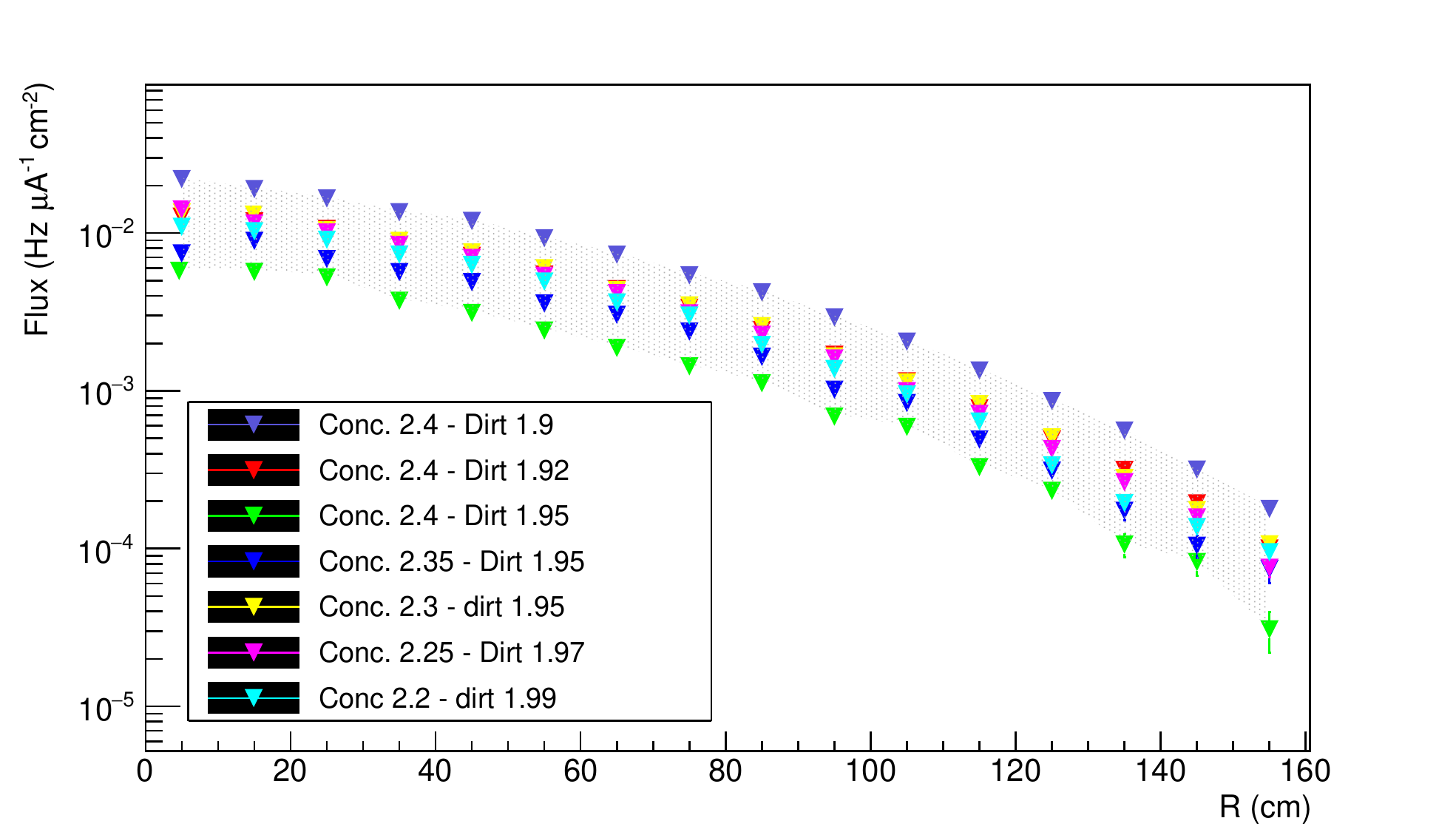}  
\caption{Simulated muon flux as a function of the sampling distance from the beam-height in Well-1 (left) and Well-2 (right). Sets of points correspond to different combination of $\rho_{dirt}$ and $\rho_{concrete}$. Values quoted in the legend are expressed in $g/cm^3$.
}
\label{fig:dirt}
\end{figure}
Simulations show a significant dependence on the dirt/concrete density. Figure~\ref{fig:dirt} show that for a mere variation of $3\%$ on $\rho_{dirt}$ around 1.93 g/cm$^3$ and $10\%$ on $\rho_{concrete}$ around 2.3 g/cm$^3$,  rates in Well-1 and Well-2 vary by 30$\%$ and $100\%$, respectively. While the absolute value is significantly affected, the shape of the distribution is much less sensitive to the density variation.
Detailed knowledge  of the dirt/concrete density/uniformity along the muon flight path is beyond the scope of this work and therefore, to compare to the data,  we run simulation with  the nominal value of $\rho_{dirt}$ = 1.93 g/cm$^3$ and $\rho_{concrete}$ = 2.3 g/cm$^3$, quoting the variation reported above as a systematic error band.

\subsubsection{Simulation results}
\begin{figure}[tph] 
\center  
\includegraphics[width=7.8cm]{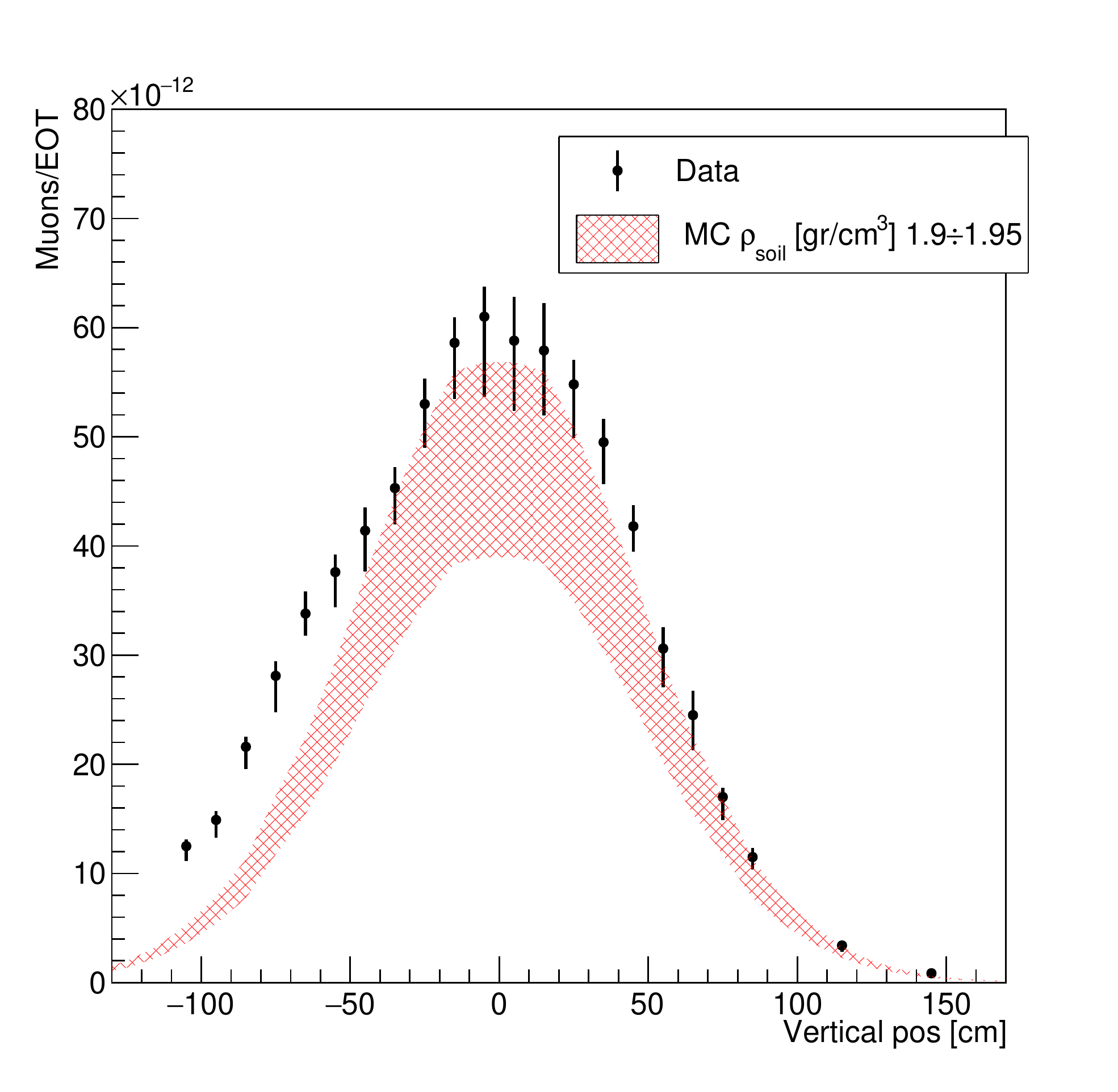}  
\includegraphics[width=7.8cm]{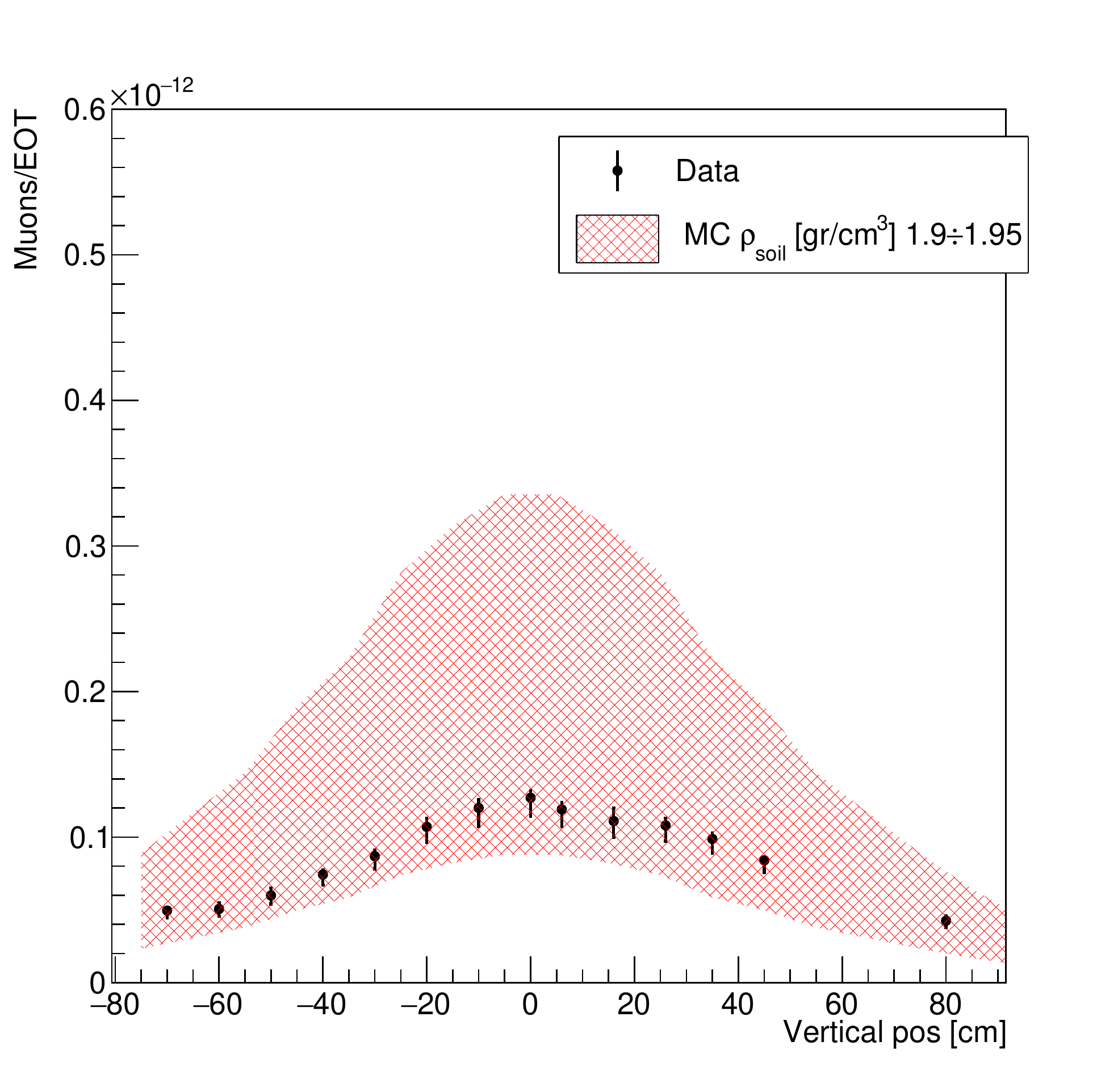}  
\caption{Comparison between simulated  and measured  muon rates. Well-1 (well-2) is  shown in the left (right) plot The red error band include the systematic error related to the density uncertainty as explained in the text.
}
\label{fig:simcomp}
\end{figure}

Figure~\ref{fig:simcomp} shows the comparison of the measured rate profiles (as a function of the vertical height) with simulations. Simulations, assuming the same test beam current, agreed very well to experimental data both on absolute values and shape of the rate profiles. Remarkably, they are able to reproduce the suppression factor of $\sim 500$ between rates measured in Well-1 and Well-2 as well as the gaussian shape and width. 
The good agreement (within the quoted systematic error) demonstrate that the simulation framework (physics processes)  and the  experimental set up implementation (dump, vault, dirt geometry and material) can be used to realistically estimate the beam-on background in the real BDX experiment configuration. 

\subsection{Other results of the test}
Besides the measurement of muons produced in the dump and the simulation framework validation presented in  previous sections, two other valuable measurements were obtained during the test: the cosmic ray rates in the BDX-Hodo and a measurement of the beam-on background taken at E$_{beam}$=4.3 GeV, a condition where all muons are expected to range out.

\subsubsection{Cosmic rays}
\begin{figure}[ht!] 
\center 
\includegraphics[width=10.cm]{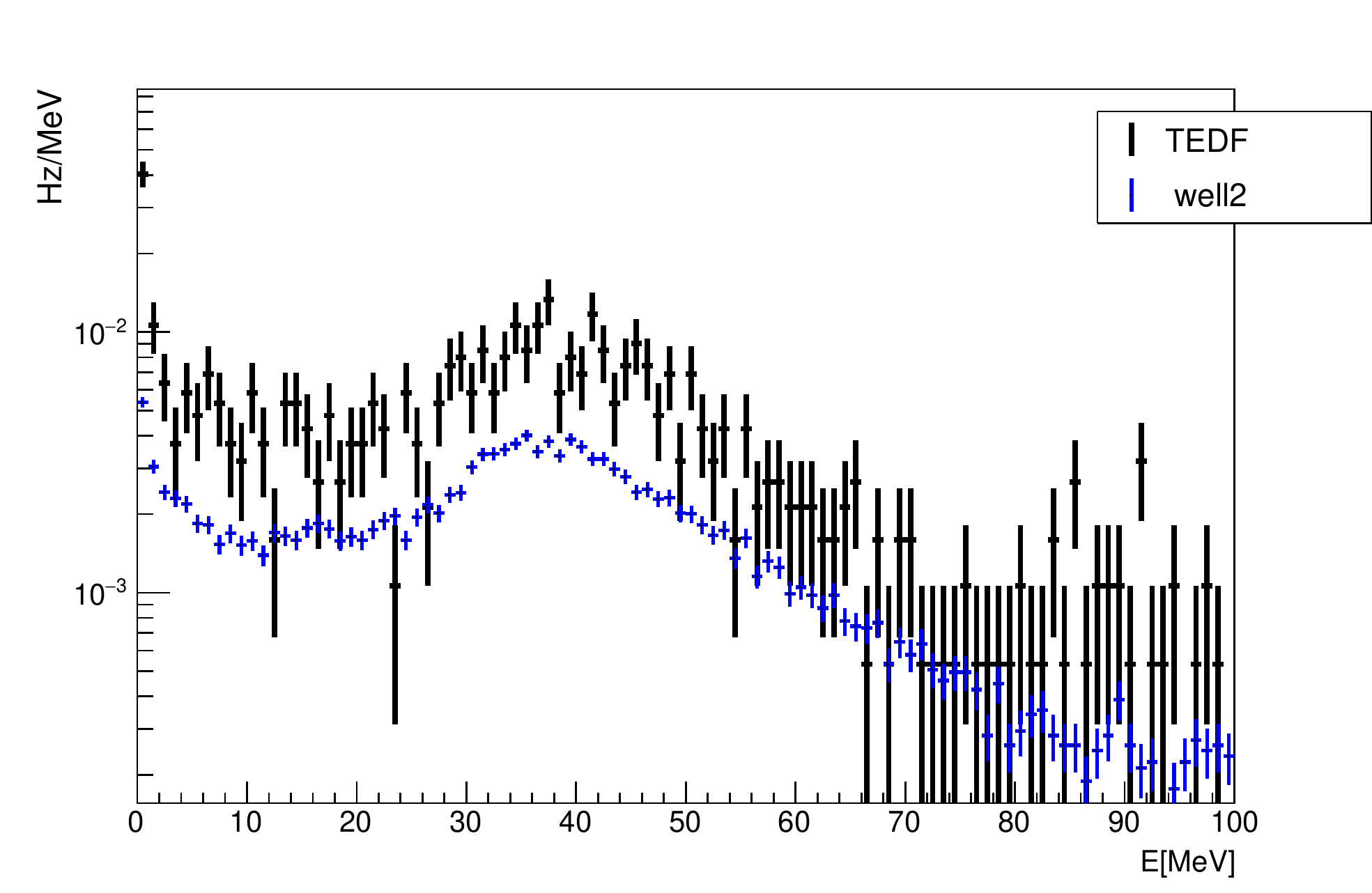}  
\caption{Rate of cosmic muons as a function of the energy deposited in the crystal when the Front/Back Paddles/Crystal three-fold coincidence is required. Blue points were taken with  BDX-Hodo in Well-2 at +30 cm. Black points were taken with  BDX-Hodo in JLab TEDF.
}
\label{fig:cosmic_3coinc}
\end{figure}

The BDX-Hodo detector was calibrated measuring cosmic muons in Genova and in the TEDF at JLab.  
Fig.~\ref{fig:cosmic_3coinc} shows the rate of cosmic events as a function of the deposited energy in the crystal applying  the same event selection used to extract the beam-related muon flux (Front/Back paddles/Crystal three-fold coincidence). 
Black points show the rate measured in the TEDF where only the roof shields the detector from cosmics.
Blue points are the result of a long (beam-off) run performed with the detector placed inside Well-2 at +30cm (w.r.t the beam height). The lower rate is explained by the partial overburden. 

In order to evaluate the potential impact of cosmic events on the beam-on flux measurement,  we extracted cosmic rates with the same procedure described above. The resulting rate is $\sim$ 0.3 Hz and $\sim$0.13 Hz in TEDF and within Well-2, respectively.  Such a rate is negligible when compared to measured beam-on rates (in all cases $>$10 Hz).

\subsubsection{Beam-on background at E$_{beam}$=4.3 GeV}
Tests performed in Spring 2018 were run in a  different experimental set-up with-respect-to the  proposed  BDX experiment (dirt shielding vs. iron , a single CsI(Tl) crystal vs. 800 crystals, few surrounding scintillator paddles vs. a two-layer fully hermetic veto system, limited  overburden, etc.). Nevertheless,
some inferrences can be drawn for the expected beam-on background in the final BDX configuration.

\begin{figure}[ht!] 
\center 
\includegraphics[width=10.cm]{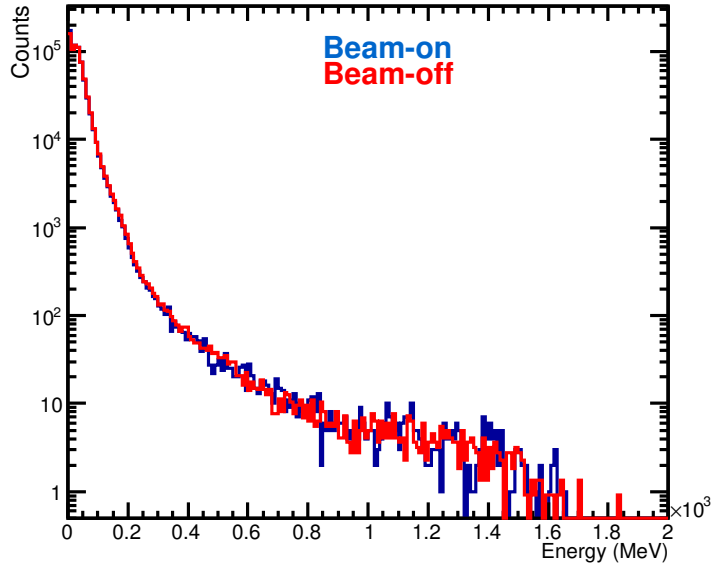}  
\caption{Blue curve: energy spectrum of the CsI(Tl) crystal located in Well-2 at the beam height,  exposed for 6.5 days to the background produced by the interaction of 4.3 GeV energy beam on the dump. Red curve: data collected in 3 weeks, and renormalized to the equivalent 6.5 days, of beam-off in the same location. The two curves (for energy larger than 100 MeV) are statistically compatible.}
\label{fig:lowenergy}
\end{figure}

In  the experimental set-up such that all muons are ranged out, simulations predict that, over a certain threshold in deposited energy in the crystal,  no counts are expected\footnote{Considering the limited exposure - about a week - and the limited sensitive mass - a single crystal -  neutrino are not detectable in this experimental configuration.}. We verified this prediction by positioning  the BDX-Hodo in Well-2 (the farthest from the dump) at beam height and exposing it to the high current ($\sim 22 \mu$A), low-energy  (4.3 GeV) Hall-A beam for  about a week (6.5 days). The accumulated charge corresponds to $\sim 7.7 \; 10^{19}$ EOT that is  remarkably close to what proposed for whole BDX experiment  ($\sim 10^{22}$). It is worth noticing that a similar EOT can not be simulated with  computing resources realistically available and therefore any information derived by this test is valuable. The corresponding energy spectrum is shown in blue in Fig.~\ref{fig:lowenergy}. In the same Figure the beam-off energy spectrum of the crystal corresponding to cosmic rays only background is also shown. The cosmics were measured in 20 days and the spectrum renormalized to the 6.5 beam-on days. The two curves overlap pretty well for all energies and, in particular for energy larger than $350$ MeV, the threshold value used to quote the BDX reach, they are statistically compatible.

These measurements confirm that no background are registered besides cosmics over a period of a week of running with 22$\mu$A continuous beam on the Hall A dump and with sufficient shielding to range out all muons. This conclusion cannot be extended directly to the full BDX experiment. However, the simulations are validated under these specific experimental conditions when Standard Model particles are properly shielded and we have collected close to 1\% of the EOT for the experiment.

\subsection{Summary}
During the spring of 2018 we measured the vertical profile of muons produced by 10.6 GeV electrons in the Hall A dump in wells dug 25 and 28 m downstream. The rates and spatial distributions of muons were reproduced by our simulation within expected uncertainties. The uncertainty in the prediction of the absolute flux is due to uncertainties in the density of dirt and concrete surrounding the dump. We also measured the energy spectrum in our detector for a week when 4.3 GeV electrons were impinging on the Hall A beam dump with an average current of 22 $\mu$A. At this incident energy, all muons range out in the dirt before reaching either well. We found no backgrounds above cosmics with energies greater than 100 MeV. 

%% file: BDX-PAC46-opt.tex
\section{BDX set-up optimization }
\label{sec:opt}
\subsection{MC simulation with FLUKA}
In this Section we report the results obtained with FLUKA~\cite{BOHLEN2014211,Ferrari:2005zk} for the beam-on background expected in the BDX experimental configuration. In particular, we focus on the results of the massive Montecarlo simulations we performed, fully exploiting biasing techniques available in FLUKA. We also discuss briefly the different shielding configurations we considered, in the process of optimizing the experimental setup\footnote{We underline that, for the different configurations we tested, muon fluxes on the detector were always smaller than $\simeq 2 \cdot 10^{-16}$ muons/EOT (this is the background value we obtained from the first run, with an un-optimized shielding), while the flux due to other particles was always found to be zero within the simulated statistics. Therefore, the optimization process required the simulation of a large number of primaries, of about $10^{16}$, for each shielding configuration being tested. This was possible only thanks to the advanced biasing features allowed by FLUKA.}.
Results confirm what was already reported in Sec.~4.3 of PR12-16-001:  \textit{provided enough shielding is installed between the beam-dump and the detector, neutrinos are the only source of beam-related background - considering a detection threshold of $O$(300) MeV.}

\begin{figure}[t] 
\center
\includegraphics[width=.8\textwidth]{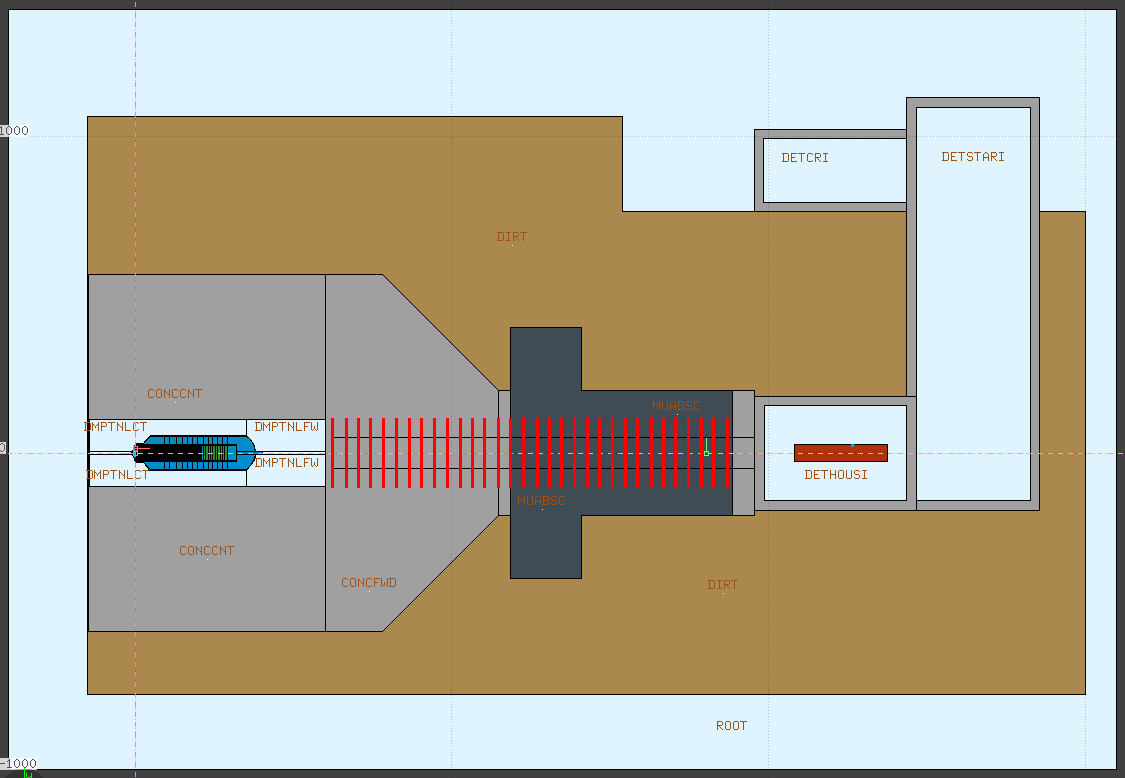} 
\caption{BDX underground facility in FLUKA, in the optimal shielding configuration. Light gray is concrete, dark-gray is the iron shielding. Red lines represent different depths in the shielding where the muon flux was scored in order to evaluate $N_{equiv.}^{EOT}$.}
\label{fig:flukageo}
\end{figure}

Starting from the current configuration of  the Hall-A beam dump geometry and materials
 implemented in FLUKA by the Jefferson Lab Radiation Control Department~\cite{jnote-bd}, we added the iron shielding and the other components of the BDX facility. Figure~\ref{fig:flukageo} shows the BDX setup implementation in FLUKA. The input card used to run the program  includes all physics processes and a tuned set of biasing weights to speed up the running time while preserving accuracy. 
We simulated an 11 GeV electron-beam interacting with the beam-dump, propagated all particles to the location of interest sampling the flux in different locations. By writing an ad-hoc routine linked to the main FLUKA executable, we also recorded each particle impinging on the detector, including its kinematics and statistical weight.

\subsubsection{High statistic simulation}

Biasing techniques available in Fluka were used in the simulation in order order to obtain the highest Montecarlo statistics with the available computing resources. By ``biasing'' we denote a set of techniques that, artificially modifying the physics model being used in the simulation, minimize the statistical fluctuations of scored quantities in a given region of interest (including both the energy range and the physical volume), while possibly increasing those elsewhere. In the following, we briefly mention the most relevant techniques we adopted, and the corresponding bias values used in the final configuration (refer to Ref.\,\cite{Ferrari:898301} for an overview of biasing techniques in Montecarlo simulations):
\begin{itemize}
\item The cross-section for the process $\gamma \rightarrow \mu^+ \mu^-$, responsible for the production of high-energy muons in the dump, was artificially enhanced by $10^5$.
\item The cross-section for photon-induced hadronic reactions was artificially enhanced by $10^2$.
\item The ``leading particle bias'' was activated for electromagnetic processes: in the electromagnetic cascade happening in the beam-dump, only one secondary particle per interaction is produced. This decreases the number of total secondaries in the shower to be tracked by the simulation.
\item Importance-sampling by splitting was implemented: the simulation volume was split in different regions of increasing importance from the dump to the detector. Each time a particle crosses a boundary, it is split in many identical copies - each with reduced statistical weight. This maximizes the number of muons penetrating deeply in the shielding, toward the detector.
\item Particle transport threshold was fixed to 100 MeV (kinetic energy)\footnote{This doesn't apply to neutrinos, for which a 10 MeV threshold was applied.}. 
\end{itemize}

The choice of importance-sampling regions, of biasing values, and of production/transport thresholds was finalized after many different trials, with the goal of minimizing the quantity $\sigma^2 \cdot T$, with  $T$ being the computation time and $\sigma$ the fluctuation of a specific flux scorer - we used the number of muons at the beam-dump exit. Results were cross-checked with those from an analogue simulation (not including any bias), with good agreement.

After finding an optimized shielding configuration for the experiment (see next Section), the corresponding high-statistics simulation was performed using the Genova cluster farm, employing $\simeq $ 300 cores for a period of $\simeq$ 3 months. A total of $\simeq $ 400k independent runs were performed, each with $5\cdot 10^5$ primary electrons, for a total of $N_0=2\cdot 10^{11}$ EOT. Being the simulation highly biased, \textit{the equivalent number of electrons $N^{EOT}_{equiv.}$ is much larger}. 

\begin{figure}[t] 
\center
\includegraphics[width=.485\textwidth]{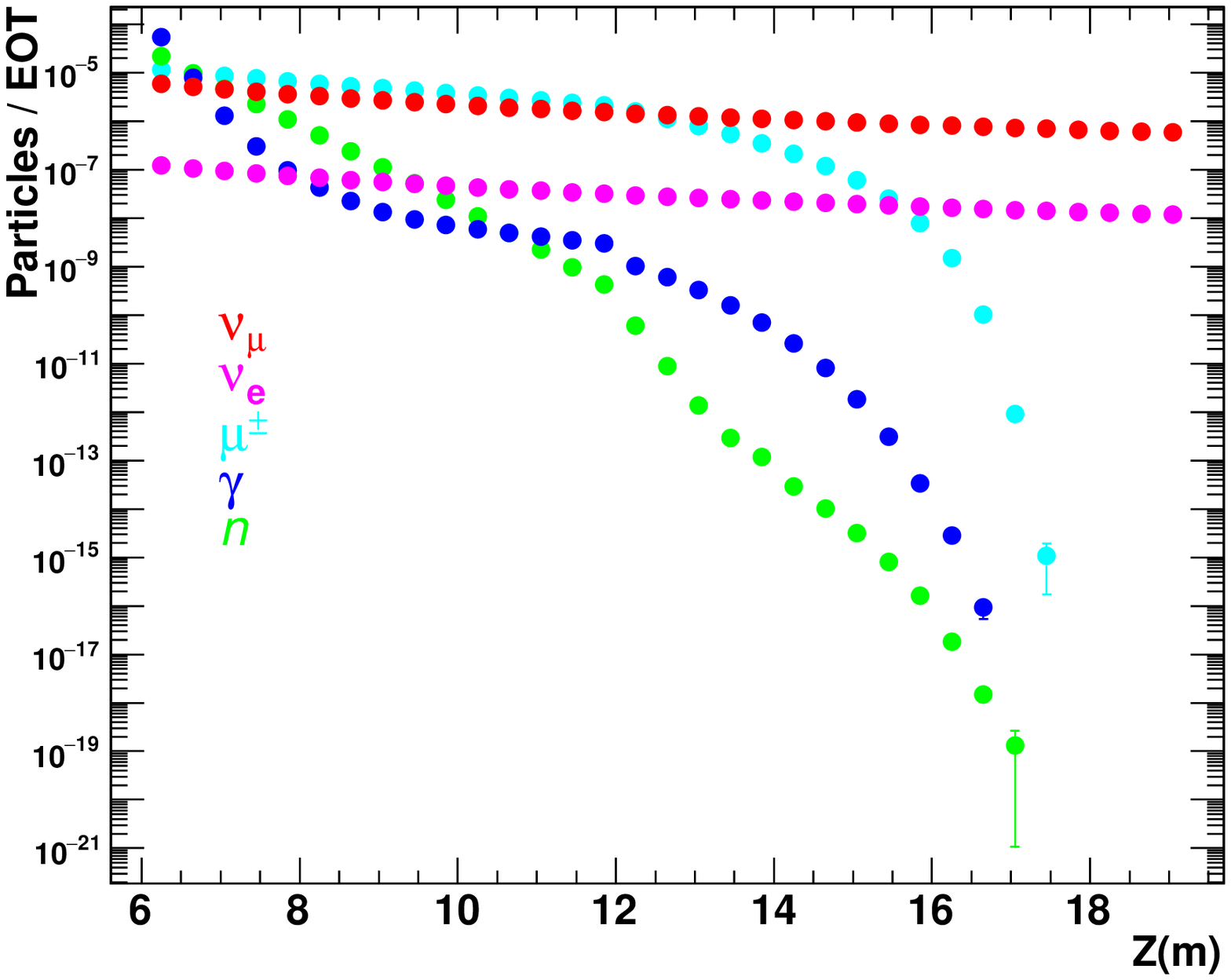} 
\includegraphics[width=.485\textwidth]{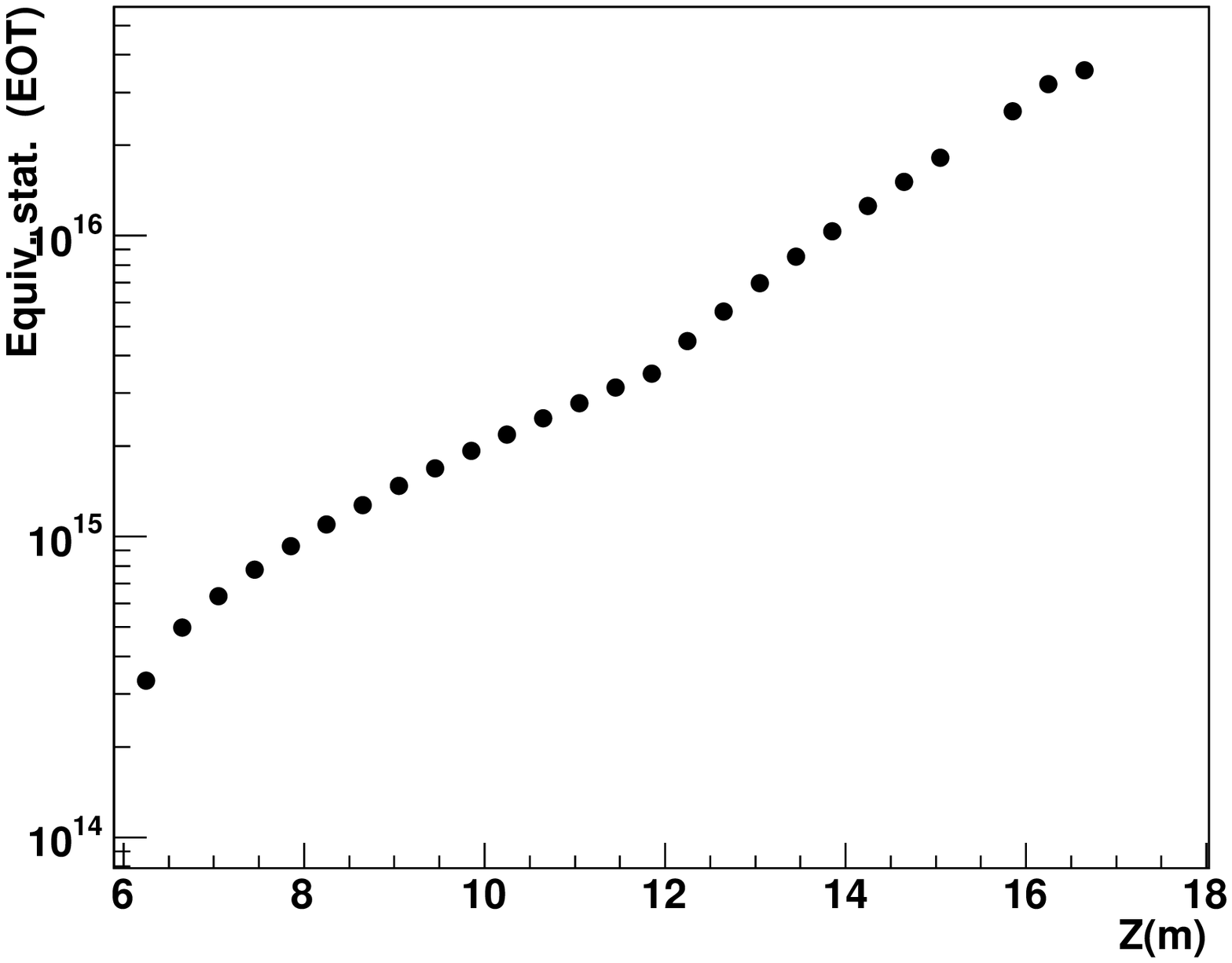} 
\caption{Left: particles fluxes per EOT at different depths in the shielding. Right: the equivalent number of electrons $N^{EOT}_{equiv.}$ in the biased simulation, as a function of the depth in the shielding.
\label{fig:eqstat}}
\end{figure}

Figure~\ref{fig:eqstat}, right panel, shows the equivalent statics obtained in the final simulation run. Each point corresponds to a different depth in the shielding were the muon flux is sampled (see also Fig.~\ref{fig:flukageo}). At larger depths in the dump, an equivalent statistics of $\simeq 0.5 \cdot 10^{17}$ EOT is obtained, to be compared to the actual number of simulated electrons $N_0=2\cdot10^{11}$. For depths $>$ 17 m, no muons have been sampled in the shielding, hence it is not possible to define $N^{EOT}_{equiv.}$. 
\begin{table}[h!]
\centering
\begin{tabular}{|c|c|c|c|}
\hline
& FLUKA & Geant4 & $N_{equiv.}^{EOT}$ \\
\hline
PAC 44 & N/A & $10^{9}$ & N/A\\
\hline
PAC 45 & $5\cdot 10^{8}$ & $5 \cdot 10^{9}$ & N/A \\
\hline
PAC 46 & $2\cdot 10^{11}$ & N/A & $5\cdot 10^{17}$\\
\hline
\end{tabular}
\caption{\label{tab:stat} Summary of statistics accumulated in different Montecarlo simulations for BDX beam-related background studies. The simulations correspond respectively to the original BDX proposal, to the PAC 45 update, and to this PAC update - the latter referring to the optimized shielding configuration.}
\end{table}
The comparison of different statistics (in EOT) simulated for the original BDX proposal~\cite{bdx-proposal}, for the PAC45 update, and for this update, is shown in Tab.~\ref{tab:stat}. We underline that, thanks to biasing, the equivalent FLUKA statistics $N_{equiv.}^{EOT}$ is much higher for about a factor $10^6$.
Thanks to the increased number of generated events and sophisticated statistical techniques, it was possible to study non-neutrino beam-related backgrounds due to very rare effects, as described in the following.

\subsubsection{Shielding optimization}

The first shielding configuration tested (Configuration A) in the optimization procedure is the one originally proposed in ~\cite{bdx-proposal} (see also Fig. 7 of the PAC45 update). Thanks to the enhanced statistics ($\simeq 10^{15} N^{EOT}_{equiv.}$ for this run), the following results were obtained:
\begin{itemize}
\item No neutrons or photons above 100 MeV transport threshold hit the detector.
\item All the muons emitted forward and passing through the shielding are ranged-out.
\item Muons emitted at a large angle in the dump, propagating in the dirt, and then, after a hard interaction, re-scattering in the detector result to a non-zero background rate. All muons have a kinetic energy lower than 300 MeV\footnote{Given the reduced statistics, it was not possible to evaluate the corresponding energy spectrum.}.
\end{itemize}
The background rate due to the latter effect was found to be of $\simeq 22$ mHz/$\mu$A, 1.5 Hz $@$ 65 $\mu$A, considering all muons above 100 MeV transport threshold. This result in $35\cdot10^{6}$ muons hitting the detector in the 285 run days corresponding to $10^{22}$ EOT.

In order to evaluate any possible contribution to the background yield, the rejection capabilities of the BDX detector should be considered. These are a combination of energy threshold and the use of the dual active-veto layer:
\begin{itemize}
\item The efficiency $\varepsilon$ of the BDX veto system to muons was measured with the BDX prototype, selecting passing-trough muons and checking events with no activity in one of the veto layers. The inefficiency $1-\varepsilon$ was found to be $1.6\cdot10^{-4}$ for the inner veto top-layer, $2.0\cdot10^{-4}$ for the inner veto bottom-layer, and $1.5\cdot10^{-4}$ for the external veto top-layer. Due to the reduced statistics, it was not possible to measure the combined inefficiency of the two veto systems. Therefore, in the following we will assume an inefficiency of $2\cdot10^{-4}$ for the whole system. \textit{This is a very conservative assumption, since the overall detection efficiency of the dual system is expected to be much better than that of each layer, close to the product of the two.}
\item Since in the simulation no muons with kinetic energy greater than the detection threshold, $\simeq 350$ MeV, were found, the effect of the latter was conservatively evaluated assuming that all muons have actually a much larger energy. Being MIP particles, the corresponding energy deposition in each crystal follows a Landau distribution. The most-probable value (MPV) was obtained from the MPV value measured with the prototype, 30 MeV. This was scaled by a factor x6 to account for the different path-length in the crystal, resulting to MPV$\simeq 180$ MeV. The probability for a muon to deposit more energy than 350 MeV$\simeq 2\times MPV$ MeV was also obtained from the data measured with the prototype and found to be $\simeq 2\%$.
\end{itemize}

The expected number of background counts is thus $\simeq140$. As discussed before, muons contributing to the background are those emitted at large angle and propagating through the dirt. Therefore, to suppress it different shielding configurations were tested, with different transverse sizes, in order to suppress this background source to $O(1)$ event.

\begin{table}[h!]
\centering
\begin{tabular}{|c|c|c|c|c|}
\hline
\textbf{Configuration} & \textbf{$Fe$ volume (m$^3$)} & \textbf{N. blocks} & \textbf{Muon rate $@$ 65 $\mu$A} &\textbf{Background counts $@10^{22}$ EOT} \\
   &   & & & (Passing all Selection Cuts) \\
\hline
A & 48 & 84 & $1.43$ Hz & 140\\
\hline
B & 109 & 190 &$0.065$  &6.4\\
\hline
C & 235 & 410 &$0.02$ &2\\
\hline
\end{tabular}
\caption{\label{tab:blocks} Shielding configuration that have been simulated during the optimization process.}
\end{table}

Results are summarized in Table~\ref{tab:blocks}. The final configuration C that was tested (see Fig.~\ref{fig:flukageo}) resulted in a rate of beam-related muons of $\simeq 0.3$ mHz/$\mu$A, corresponding to 1 background event in the full BDX run, with the very conservative hypothesis made to estimate it. It is worth mentioning that, with this shielding configuration, the rate of beam-related muons that hit the detector (0.02 Hz), is $\simeq$ 1000 times lower than the cosmogenic background. From this study, we conclude that \textit{we have identified an optimal shielding configuration, resulting in a completely negligible non-neutrino beam-related background.}

\subsection{The neutrino background}

Neutrinos ($\nu_{e}$, $\nu_{\bar e}$,  $\nu_{\mu}$, and   $\nu_{\bar \mu}$) are produced in muon decays and hadronic showers (pion decay). The majority come from pion and muon decay at rest but a non negligible fraction, due to in-flight pion decay, experience  a significant boost to several GeV energy.
High energy neutrinos interacting with BDX detector by elastic and inelastic scattering may result in a significant energy deposition -$O(300)$ MeV- that may mimic an EM shower produced by the $\chi$-atomic electron interaction.

In order to evaluate the background contribution due to neutrino interactions, we employed the following multi-step procedure:
\begin{itemize}
\item The differential neutrino flux, with respect to energy and angle, was sampled on the front-wall defining the underground hall where the detector is located. To perform this calculation, results from the FLUKA high-statistics simulation previously described were used.
\item Neutrinos were propagated from the front-wall to the detector volume, where an interaction with the $Cs$/$I$ nuclei was forced. The primary vertex was randomly distributed along the neutrino trajectory within the detector volume. Both the total interaction cross-section (averaged over the two nuclei) and the kinematics and topology of produced particles were sampled on an event-by-event basis.
\item The response of the detector to neutrino secondaries was included in the simulation.
\end{itemize}
This procedure allowed as to compute the neutrino flux once.  The second and third steps were repeated as needed to study different detector positions and setups.

The first step of the calculation, i.e. the evaluation of the neutrino flux, was performed by considering a simplified model of the Hall-A beamline, including only the beam-dump, and assuming a pencil-like 11-GeV beam. The latter assumption is conservative, since an angular spread in the primary beam would be reflected in the produced $\nu$s, reducing the actual flux on the detector. Possible effects due to the first approximation were evaluated by comparing present results to the flux obtained considering following configurations: a realistic description of the Hall-A beamline, including the $50\%X_0$ diffuser, and the configuration specific to the Moller experiment configuration~\cite{moller}. Results are discussed in Appendix~\ref{appx:moller}. \textit{No significant effects affecting the final results were found.}

In the second step, neutrino-nucleus interactions were simulated by using NUNDIS and NUNRES~\cite{Battistoni:2009jen,Battistoni:2009zzb}, the FLUKA internal neutrino-nucleon interaction generators. NUNDIS/NUNRES were developed on the basis of the results achieved by the effective nuclear models implemented in FLUKA, which have good predictive power in hadron interactions. 
The codes simulate both charged current (CC) and neutral current (NC) interactions, for all neutrino species. NUNDIS and NUNRES are extensively used in the neutrino community to generate neutrino-induced events. To name few applications, the two codes have been used in the simulation of~\cite{2009AIPC.1189..343B}:
\begin{itemize}
\item  Events  in ICARUS  from  the CNGS beam ($E \simeq 18$ GeV); 
\item  Atmospheric neutrino interactions  in ICARUS  (with good identification  of events); 
\item Search  for  sterile neutrinos  at CERN-PS  in the framework  of a new project DOUBLE-LAr ($E<2$ GeV)
\item Measurements  of the $\theta_{13}$ in the framework  of a new project MODULAr in the off-axis CNGS beam. 
\end{itemize}

In order to further validate the results, we compared them with those obtained using the GENIE code~\cite{genie}, also widely used in the neutrino community.  The comparison is discussed in Appendix~\ref{appx:genie}. Results obtained from the two codes were found to be in very good agreement.


\subsubsection{Neutrino flux}

\begin{figure}[t] 
\center
\includegraphics[width=.7\textwidth]{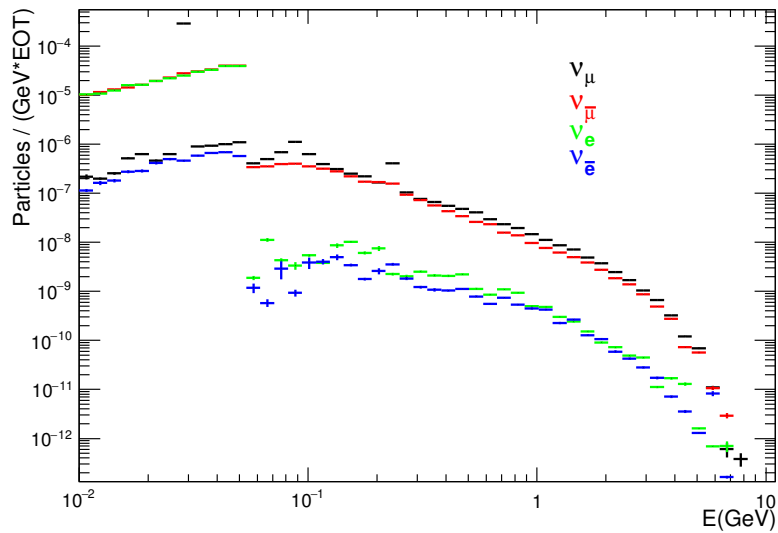} 
\caption{Differential energy spectrum of neutrinos ($\nu_{\mu}$, $\bar\nu_{\mu}$, $\nu_{e}$,  and $\bar\nu_{e}$)  impinging on the  BDX detector volume.}
\label{fig:neutflux-det}
\end{figure}

A sizable number of neutrinos propagate to the BDX detector.  
Figure~\ref{fig:neutflux-det} shows the differential neutrino spectra sampled at the detector front face. A tiny but not negligible part of the spectrum has energy greater than the detection threshold, $O$(300) MeV (see next Section).
These events may produce signals in the BDX detector that mimic lDM interactions, thus generating backgrounds and limiting the experimental reach.

\subsubsection{Neutrino interactions in the detector}

As previously described, the FLUKA code was used to simulate neutrino interactions in the detector volume. For each impinging neutrino, both the total interaction cross-section and the kinematics of secondary particles from the reaction were saved on an event-by-event basis, to allow the subsequent simulation of the detector response\footnote{FLUKA only considers $\nu$-N and $\bar\nu$-N interactions disregarding  $\nu$-electron and $\bar\nu$-electron interaction.
We checked the validity of this approximation doing an analytical estimate of the $\nu$-e and $\bar\nu$-e contribution finding that for 10$^{22}$ EOT we are expecting less than 1 interaction in the BDX volume, for an energy threshold of $O$(100) MeV.}. To speed-up calculation, only $\nu$s with energy greater than 100 MeV were considered - the cut-off still being lower than the foreseen detection threshold (see next Section).

\begin{figure}[t] 
\center
\includegraphics[width=.45\textwidth]{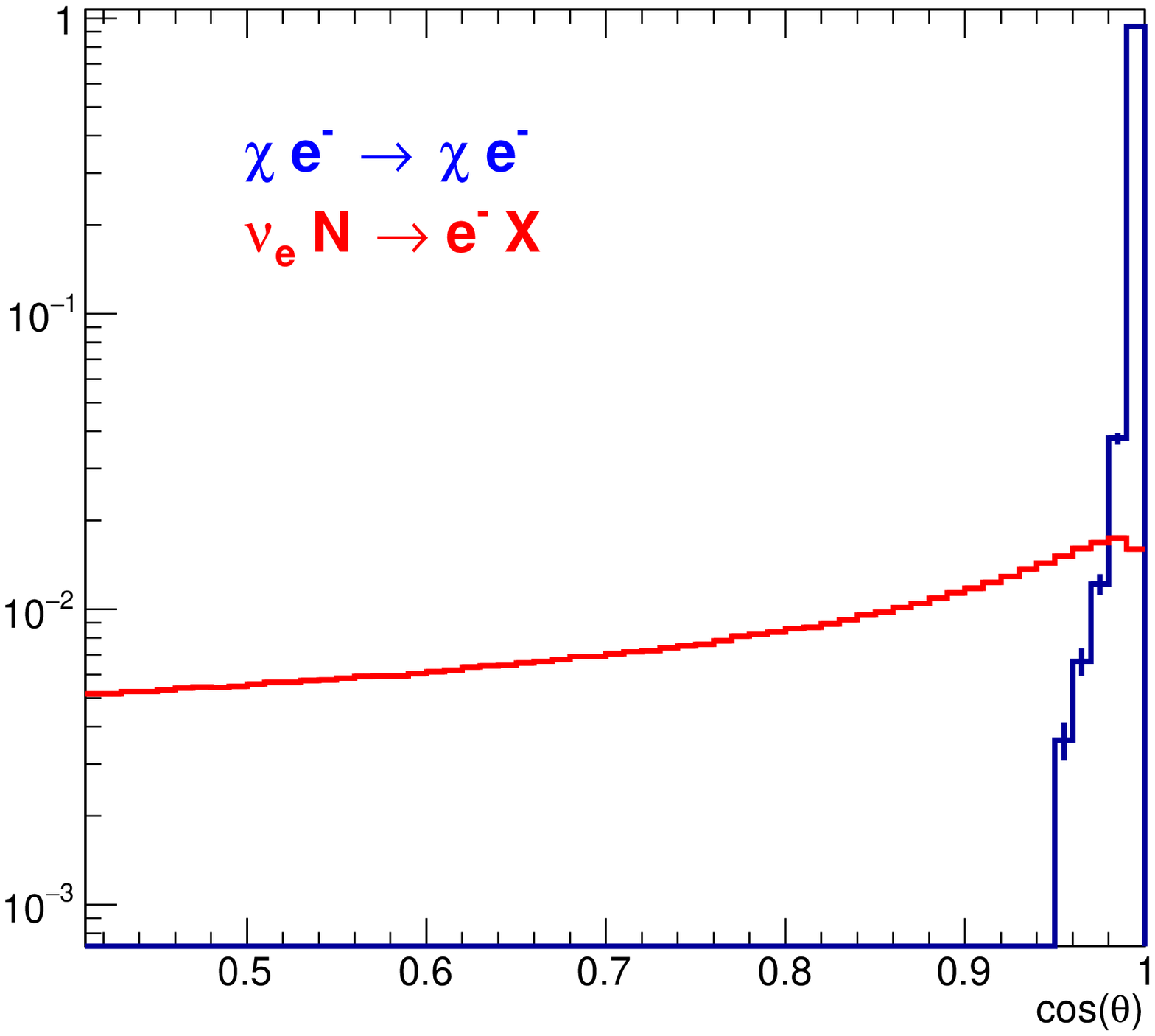} 
\includegraphics[width=.45\textwidth]{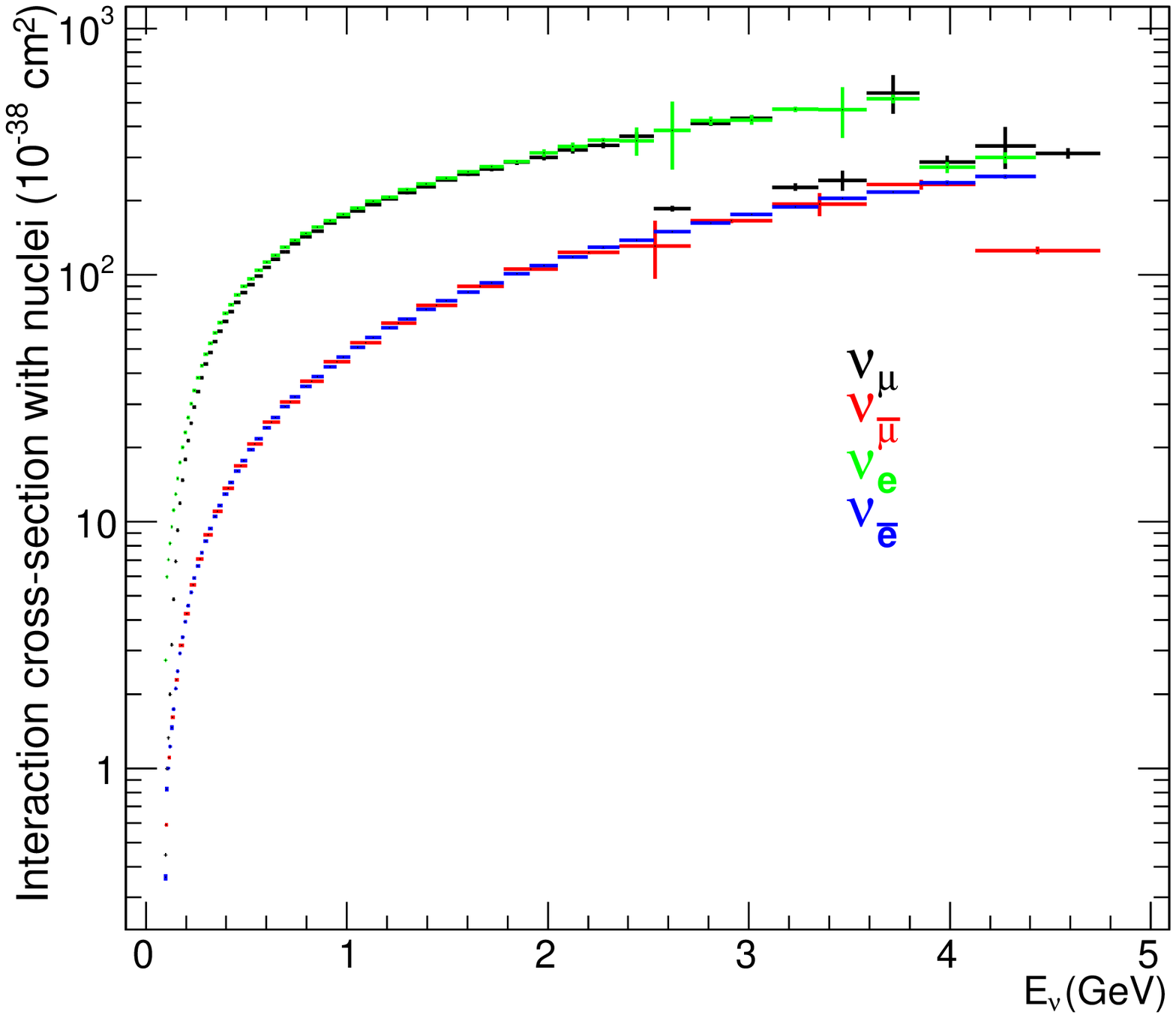} 
\caption{\label{fig:nu1} Left: scattered electron angle distribution for the signal ($e^{-}\chi \rightarrow e^{-}\chi$) and $\nu_e$ CC background ($\nu_e N \rightarrow e^- X$) reactions. The two histograms have been scaled to the same unitary area. Right: the total $\nu-N$ interaction cross-section for $CsI$ (averaged over the two nuclei).
\label{fig:fluka-bd}
}
\end{figure}

The implication  for the $\chi$-electron signal measurement for different $\nu$-matter interactions are listed below.
\begin{itemize}
\item{$\nu_e N \to e  X$: \textbf{this is the most critical background source for the experiment}, since the CC interaction could produce a high energy electron into the detector that mimics the signal. This background can be rejected considering again the different kinematics of the $\nu$ interaction with respect to the  $\chi$-electron scattering. The significant difference in the polar angle of the scattered electron(with respect to the beam direction) allows to define a selection cut to identify $\nu_e$ and separate from the $\chi$. This difference is shown in Figure~\ref{fig:nu1} (left panel), reporting the angular distribution of scattered $e^-$ from $\nu_e$ CC, compared to the characteristics kinematics of the $\chi e^-\rightarrow \chi e^-$ kinematics.}
\item{$\nu_\mu N \to \mu X$: the CC interaction produces a $\mu$ in the final state (beside the hadronic state $X$). This reaction can be identified and used to provide an experimental assessment of the $\nu_\mu$ background (and therefore estimate the $\nu_e$ contribution) by detecting a $\mu$ scattering in the detector (a MIP signal inside the calorimeter with or w/o activity in IV and OV) or, alternatively, selecting kinematics in which the $\mu$ is emitted at large angles.}
\item{$\nu_\mu N \to \nu_\mu X$: the NC interaction produces an hadronic state $X$ that may interact in the detector (while the scattered $\nu$ escapes from detection). This can mimic an EM shower if $\pi^0$ ($\gamma$'s) are produced. However, due to the difference in mass, the scattered $\nu$ carries most  of the available energy providing a small transfer to the hadronic system and reducing the probability of an over-threshold energy deposition.}
\item{$\nu_e N \to \nu_e X$: same considerations as above.}
\end{itemize}
We underline that, in the simulation, \textit{all the $\nu$ interaction mechanisms have been included,}  thus accounting for their possible contribution to the total background yield.

\subsubsection{Neutrino-induced background events}

The number of neutrino-induced background events in the detector depends on the choice of the selection cuts. These have to be tuned by coherently looking at the effect on the signal efficiency and on the background rejection, as discussed in details in the next Section. For the optimized experimental configuration there described, the number of foreseen neutrino events is $\simeq$ 5.

%% file: BDX-PAC46-opt2.tex
\subsection{Detector optimization and reach}

In this section we report the results of the optimization process performed on the BDX detector  and signal  selection cuts. As described in appendix A, the sensitivity $s$ of a counting experiment  depends on the foreseen average of  background counts $B$:  
\begin{equation}
s \simeq 2.3 + 1.5 \sqrt{B}.
\end{equation}
In the case of BDX, the expected signal counts $N$ (at a given value of the $\chi$ mass) depends on the coupling $\varepsilon$; thus, the minimum value of $\varepsilon$ that BDX can probe is given by:
\begin{equation}
N(\varepsilon_{min},m_\chi) = 2.3 + 1.5 \sqrt{B},
\label{eq:reach}
\end{equation}
where $N$ and $B$ depend on the detector setup as well as on the cuts adopted. The optimization performed consists in the identification of the detector-cuts configuration that results in the minimum achievable $\varepsilon_{min}$ in the largest possible $m_\chi$ range. It is worth mentioning that statistical procedure allows to handle coherently the signal and the background, and the effect on these of the selection cuts. In the following, we assumed that $B$ is known with a negligible uncertainty. In any case, considering a possible uncertainty on the number of backgrounds $\sigma_B\simeq O(B)$, and given the number of expected background in the optimized configuration of $\simeq 5$, the correction to the sensitivity is $O(1)$, and thus negligible. In the BDX experimental analysis, these effects will be properly accounted for through \textit{nuisance} parameters, as described in App.~\ref{app:s}.

\subsubsection{Signal and background description}

The theoretical scenario addressed here is the minimal dark photon model: $A'$ are produced in the beam-dump via $A'${\it{-strahlung}} and decay in a $\chi\bar{\chi}$ pair; $\chi$ particles interact in the detector volume via elastic scattering with electrons. 
The study of the response of the BDX detector to $\chi - e$ scattering has been performed through GEMC simulations.  For a set of $m_\chi$ values in the range 1-150 MeV, a custom generator was used to generate events within the BDX detector volume, considering the interaction process with atomic electrons. The generator provided the total number of foreseen signals $N_{I}$ in the detector per EOT, at a  specific value of the coupling $\varepsilon_{ref}$. Since $N_{I}$ scales as $\varepsilon^4$ the number of interaction for an arbitrary $\varepsilon$ value is: 
\begin{equation}
N_{I}(\varepsilon)= \frac{\varepsilon^4}{\varepsilon_{ref}^4}N_{I}
(\varepsilon_{ref})
\end{equation}
As for the signal, neutrino background characterization was performed through  simulations. For a detailed description of the simulation chain, see the previous section.

The projections for cosmic background events have been obtained from the measurement of the BDX prototype detector at LNS Catania~\cite{bdx-proposal}. The prototype, a single CsI(Tl) crystal enclosed into  two scintillating veto layers and a lead shielding, measured cosmic background in a configuration similar to that proposed for the BDX experiment. The expected number of cosmic  events as a function of the energy threshold was extrapolated from the spectrum of events in anti-coincidence with both vetos measured with the prototype (see Tab.~\ref{tab:cosm_th}).
\begin{table}[h!]
\centering
\begin{tabular}{|c|c|}
\hline
Energy Thresold (MeV) & Expected Counts (285 days meas.) \\
\hline
200 & $740 \pm 300$\\
\hline
250 & $57 \pm 25$\\
\hline
300 & $4.7 \pm 2.2$\\
\hline
350 & $0.037 \pm 0.022$\\
\hline
\end{tabular}
\caption{\label{tab:cosm_th} Number of expected cosmic background events as a function of the single crystal energy threshold.}
\end{table}

\subsubsection{Detector and cuts optimization}

\begin{figure}[t] 
\center
\includegraphics[width=14.5cm]{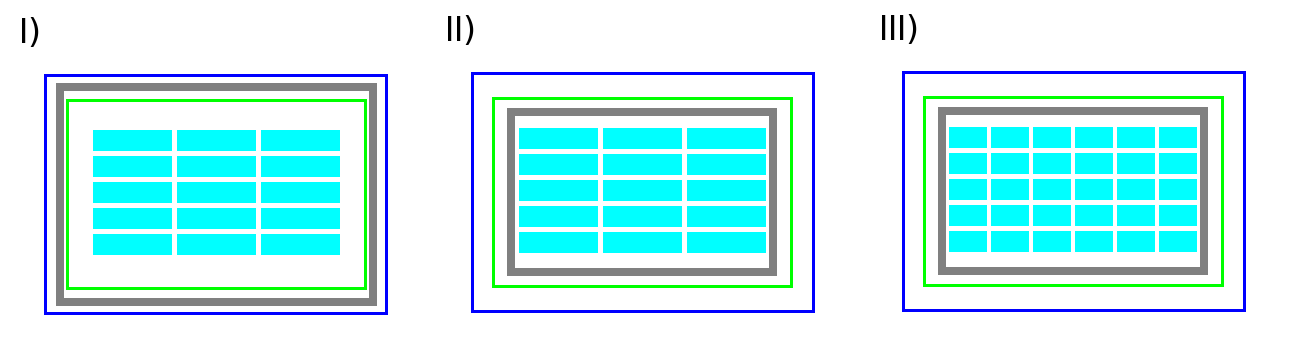}
\caption{Schematic of the three detector configurations tested (lateral view): I) nominal, II) inner lead, III) inner lead, half length crystals. Crystal are drawn in cyan, internal veto in green, external veto in blue and lead in gray.}
\label{fig:d_scheme}
\end{figure}

The original concept of the BDX detector~\cite{bdx-proposal} features an electromagnetic calorimeter made of $\sim 800$ CsI(Tl) crystals ($\sim 5\times 5\times 30\,\, cm^3$ each), for a total volume of order $1\sim m^3$. The calorimeter is enclosed within two active veto layers made of plastic scintillators and a 5 cm thick lead layer placed between the two vetos. To optimize this design we tested two new configurations (other than the nominal one), slightly varying the components arrangement (see Fig.~\ref{fig:d_scheme}), without changing the total active volume foreseen in the BDX proposal~\cite{bdx-proposal}. 
For the first alternative setup tested, the lead shielding was moved inside the internal veto layer. This setup is motivated by the observation that, in the nominal configuration, electromagnetic shower produced by $\chi-e$ interaction may hit the internal veto resulting in a low signal efficiency. The second variation consisted in reducing the crystals dimensions by a factor 2 (from 30 cm to 15 cm length). This test was performed to check if increasing the calorimeter segmentation provides higher background rejection capability. 
For each detector variation, signal and neutrino background were simulated, and events were reconstructed and analyzed applying different set of selection cuts. In particular, we investigated the effect of cuts on four different measurable quantities:
\begin{itemize}
\item \textbf{Seed energy $E_{seed}$}: the highest energy measured in a single crystal within one module\footnote{A module is defined as a $10 \times 10$ crystals matrix of the calorimeter, arranged perpendicularly to the beam direction. In the BDX detector (nominal setup), the calorimeter is composed of 8 modules. }. 
\item \textbf{Module energy $E_{M}$} : total energy measured in one module.
\item \textbf{Number of hit $N_{hits}$}: number of crystals hit in a module per event.
\item \textbf{Shower transverse dimension $R$}: quantity indicating the shower deviation from the beam direction\footnote{The quantity $R$ is defined for a single module; if an event hits more than one module, the $R$ value used for event selection is the one of the module with the higher measured energy $E_M$}. It is defined as follows:

\begin{equation}
R^2 = \frac{\sum_{i=1}^{N_{hits}}w_i X_i^2  - (\sum_{i=1}^{N_{hits}}w_i X_i)^2}{\sum_{i=1}^{N_{hits}} w_i} + \frac{\sum_{i=1}^{N_{hits}}w_i Y_i^2  - (\sum_{i=1}^{N_{hits}}w_i Y_i)^2}{\sum_{i=1}^{N_{hits}} w_i}.
\end{equation}

Here $i$ runs over the crystals hit in the module, $X_i$ and $Y_i$ are the geometrical indexes of each crystal and $w_i$ is a weight factor accounting for the energy of the $i$-th hit $E_i$:
\begin{equation}
w_i = \max[\,0,\,3.1 + \ln(E_i /E_M)];
\end{equation}
here, the logarithmic factor prevents from overestimating the contribution of the low energy tails of the electromagnetic shower. 
\end{itemize}
In order to determine if an event  passed the selection, at least one of the modules of the calorimeter had to fulfill the cuts. The optimization of the threshold values for $E_{seed}$,${E_M}$,$N_{hits}$ and $R$ was performed using a custom code: the number of expected neutrino events $N_\nu$ and signal events $N_s(m_\chi)$ (for $\varepsilon = 1$) were estimated varying the threshold values among a large set of configurations; cosmic background events $N_c$ were extrapolated from Tab.~\ref{tab:cosm_th}. Finally, for each cuts configuration the sensitivity curve $\varepsilon_{min} (m_\chi) $ was calculated   using Eq.~\ref{eq:reach}:
\begin{equation}
 \varepsilon_{min}^4 = \frac{2.3 + 1.5 \sqrt{N_\nu + N_c}}{N_s (m_\chi)}.
\end{equation}

\subsubsection{Results}
Among all the configuration tested, the set of cuts giving the best reach, for all the detector setups, is the following:
\begin{equation}
E_{seed}>350 \,MeV; \,\,N_{hits}\geq1; \,\,R<0.6;
\end{equation}
with no prescription on $E_M$. This result reflects the fact that, for both signal and $\nu$ background, a relevant fraction of events produces a single crystal hit with high energy, which makes ineffective any cut on the hit multiplicity. It should be noticed, however, that for all events with more than one hit in a single module, the cut on the variable $R$ provides an efficient rejection of $\nu_e$ background, because of the different kinematics between $\chi$ and $\nu_e$ interactions in the detector (see previous section).   

\begin{figure}[t] 
\center
\includegraphics[width=14.5cm]{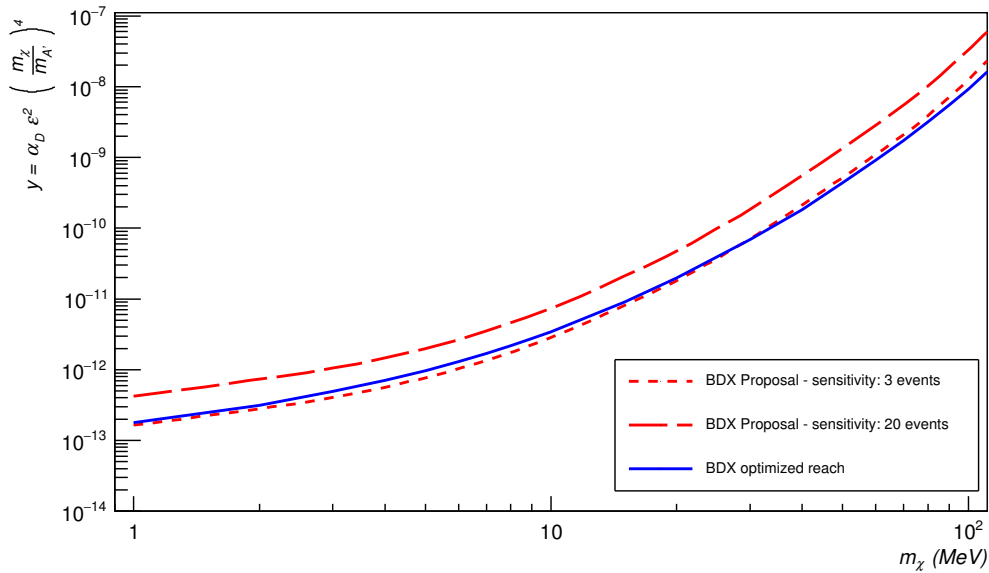}
\caption{Comparison between BDX reach with optimized cuts and detector configuration (inner lead, nominal crystals dimensions) and the reach proposed in the BDX Proposal, for an assumed sensitivity of 3 and 20 events. Here $y=\alpha_D\varepsilon^2(m_\chi / m_{A'})^4$, 
with $\alpha_D=0.5$, $m_{A'} =3 m_\chi$.} 
\label{fig:bdxreach} 
\end{figure}

Regarding the detector setup, results confirm that the arrangement with the lead shielding placed within both veto layers provides a better efficiency for signal, up to a factor 2 for large $m_\chi$ values. This translates in a clear improvement on the reach.
On the other hand, the detector variation with higher segmentation in the calorimeter (crystals length reduced by a factor 2), do not produce a relevant improvement in the experiment sensitivity. In fact, to increase significantly the background rejection capability of the detector, an even finer segmentation would be necessary, requiring a substantial increase of the number of channels of the calorimeter. Consequently, we selected the setup with internal lead shielding and full size crystals as the best option for the experiment. 

The detection efficiency for signal events in this configuration is of order of $10\%-40\%$, depending on the $A'$ mass. This value accounts for both selection cuts and edge effects: signal events happening in the outermost crystals have a lower reconstruction efficiency due to the badly contained electromagnetic shower. Nevertheless, an analysis procedure foreseeing a fiducial volume excluding outermost crystals, can always be performed for the final experiment whenever any critical issue will be found for edge events. This analysis will benefit from the full information available at single event level for each detector channel.
The number of expected background events is $B \sim 5$ (all due to neutrinos), which corresponds to a sensitivity $s$ of $\sim 6$ events for signal. 

Fig.~\ref{fig:bdxreach} shows BDX reach in this setup compared to the reach quoted in the BDX proposal~\cite{bdx-proposal}. The proposal curves had been derived  with the reasonable assumption of a $20\%$  signal efficiency and a sensitivity $s$ in the range of 3-20 events. Red dashed lines correspond to the two extremes of this interval. The reach obtained with the optimization process described in this section (blue solid line) is comparable to that quoted in the  proposal for the most favorable hypothesis  $s=3$; this proves the robustness of the assumptions made in~\cite{bdx-proposal}.

%% file: BDX-PAC46-upd-appxA.tex
\section{BDX sensitivity evaluation procedure}\label{app:s}

In this Appendix, we discuss the numerical strategy that was adopted to compute the BDX sensitivity. The input ingredients for this calculation are: the foreseen backgrounds and the expected signal yield, for a given combination of the model parameters. Both depend on the selection cuts adopted in the analysis. Therefore, it is mandatory to evaluate and optimize the sensitivity within a coherent framework that accounts simultaneously for both effects.

The \textbf{sensitivity} $s$ of the BDX experiment is defined as \textit{the average of the upper limits that would be reported by a set of equal experiments, performed within the same conditions, under the hypothesis of no signal.} For the definition and computation of the upper limit, we use the Bayesian approach - that, in case of a counting experiment (as BDX), is expected to give more conservative results than the corresponding frequentist approach. The two procedures converge to the same result in case of a small number of expected background events.
\subsection{Upper limit computation}

Consider an experiment where the average number of background events is $B$ and the foreseen number of signal events is $S$. While $S$ is not known - and the goal of the procedure is to produce a confidence interval for it, $B$ is assumed to be known without uncertainty \footnote{In case this hypothesis is not strictly verified, the procedure can be generalized, by accounting for possible uncertainties in $B$ through corresponding \textit{nuisance} parameters. As a rule of thumb, if $\sigma_B$ is the uncertainty in the average background value $B$, the sensitivity is obtained through the prescription $B\rightarrow B+\sigma^2_B$.
}.
The number of measured events $n$ is thus governed by a Poissonian distribution distribution with average value $\mu=S+B$:
\begin{equation}
P(n;S,B)=\frac{(S+B)^n e^{-S-B}}{n!} \; \; ,
\end{equation}
where  is the probability that the experiment will measure $n$ events, 

In the frequentist approach, it is only possible to introduce the above probability $P(n;S,B)$ to measure a certain value of $n$, given $S$ (unknown) and $B$ (known). From this, an upper limit can be derived, for example trough Von-Neumann confidence belts construction. In the Bayesian approach, instead, one can introduce the probability $P(S;n,B)$ that $S$ has a certain value, we can talk also given the measured value of $n$ and the value of $B$. This is done through the Bayes theorem:
\begin{equation}
P(S;n,B) = \frac{P(n;S,B)\cdot P(S)}{P(n)} \; \; ,
\end{equation}
where:
\begin{itemize}
\item $P(S)$ is the so-called "prior" distribution, i.e. the probability distribution function of $S$ \textit{independent from the measurement}. This summarizes the knowledge about $S$ before actually performing the measurement. It is a "subjective" function. In the following, we'll make the common choice of only assuming that $S$ is non-negative, i.e. $P(S)=\theta(S)$, with $\theta$ being the Heaviside step function.
\item $P(n)$ is the probability distribution function of $n$, for any value of $S$. This can be written as: $P(n)=\int dS P(n;S,B)P(S)$, where the integral is performed over all possible values of S.
\end{itemize}
Explicitly:
\begin{equation}
P(S;n,B) = \frac{P(n;S,B)\cdot \theta(S)}{\int_0^{+\infty} P(n;S,B) dS}
\end{equation}
The above equation is a PDF for $S$. The upper limit at confidence level (CL) $1-\alpha$ can thus be derived, imposing that (using standard notation):
\begin{equation}
 \int_0^{S_{up}}  dS P(S;n,B) = 1 - \alpha 
\end{equation}
and inverting this to determine $S^{\alpha}_{up}$.
Using the explicit form of $P(n;S,B)$:
\begin{align} 
\frac{ \int_0^{S^{\alpha}_{up}} dS P(n;S,B)}{\int_0^{+\infty}  dS P(n;S,B)} & = 1 - \alpha  \\
\frac{ \int_0^{S^{\alpha}_{up}} dS (S+B)^n e^{-S}}{\int_0^{+\infty} dS (S+B)^n e^{-S}} &= 1 - \alpha 
\end{align}
This equation will result in the upper limit $S_{up}$, that depends from the value of background $B$ and the number of measured counts $n$: $S_{up}(n,B)$.
\footnote{The special case of $B=0$, $n=0$ can be solved exactly. This is the case of an experiment with no expected background events, that measures $n=0$:
\begin{equation} 1-e^{S_{up}} = 1 - \alpha \; \; ,\end{equation} that for the $90\%$ upper limit gives: $ 1-e^{S_{up}} = 0.9 $, $S_{up}=2.3$.}

\subsection{Sensitivity evaluation}
Since $n$ is a stochastic variable, \textit{$S_{up}$ is a stochastic variable also.} Therefore, different equivalent experiments, performed with the exactly the same conditions, may report different values for it. 
\begin{figure}[t]
\centering
\includegraphics[width=.7\textwidth]{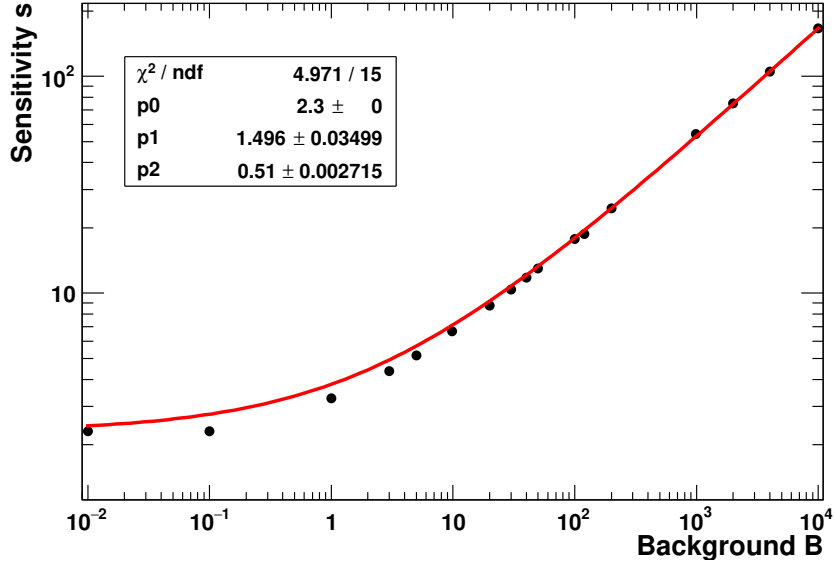}
\caption{\label{fig:sensitivity} Sensitivity $s$ for a counting experiment, at $90\%$ CL, for different values of the foreseen background $B$.}
\end{figure}
The sensitivity $s_\alpha$, i.e. the average under the hypothesis of no signal, is thus:
\begin{equation}
s_\alpha = \sum_0^{+\infty} P(n;B,S=0) S^{\alpha}_{up}(n,B)
\end{equation}
The numerical evaluation of this series can be performed through a Montecarlo technique: a large number of pseudo-experiments is generated, each with the value of pseudo-measured $n$ sampled from the PDF $P(n;S=0,B)$. The the upper limit $S_{up}$ is evaluated for each of them  through the above prescription. Finally, the average of the obtained results is computed.

The result of the calculation, at $90\%$ CL, is shown in Fig.~\ref{fig:sensitivity},
for different values of the expected background $B$. The red curve is the result of the best-fit performed with the function:
$ s = 2.3 + p_2 \cdot B^{p_1}$, where the factor 2.3 comes from above considerations about upper limits when B=0. The obtained result for the $p_1$ parameter, $p_1 \simeq 0.5$, is a consequence of the well-known behavior of the sensitivity to scale as $\sqrt{B}$ for sufficiently large values of $B$.

%% file: BDX-PAC46-upd-appxB.tex
\section{Comparison of FLUKA and GENIE results for $\nu$ interactions simulation}\label{appx:genie}

In this Appendix, we discuss selected results regarding the simulation of $\nu$ interactions with the detector nuclei. We compare results obtained with the FLUKA code (NUNDIS/NUNRES) employed in the BDX background evaluation with those obtained from the GENIE~\cite{genie} code, widely used in the neutrino community to simulate $\nu-$induced interactions on atomic nuclei. Results show a good agreement between the two codes, thus confirming the robustness of background calculations we performed.

Specifically, we compared results obtained considering the scattering of $5-$GeV $\nu_\mu$ and $\overline{\nu}_\mu$ on a $^{133}$Cs nuclei. This value is representative of the high-energy portion of the $\nu$ spectrum expected in BDX, see Fig.~\ref{fig:neutflux-det}. Also, we limited the comparison to $\nu_\mu$ and $\overline{\nu}_\mu$ since $\nu_e$ and $\overline{\nu}_e$ interactions properties are similar.
Results are summarized in Tab.~\ref{tab:comparison}, reporting the total interaction cross-section (per nucleon), and the fraction of events related to a CC interaction\footnote{The GENIE event generator tags events as being produced by NC or CC interactions. In FLUKA, on the other hand, NC-induced events were identified as those where a $\nu$ with energy $>10\%$ of the impinging $\nu$ energy was present in the final state.}. The results obtained with the two numerical codes are in a reasonable agreement. The obtained CC cross-sections ($\sigma^{CC} = \sigma \cdot CC^{frac}$) are also in excellent agreement with experimental data~\cite{Formaggio:2013kya}.

\begin{table}[h]
\centering
\begin{tabular}{|c||c|c|c|}
\hline
\textbf{Observable} & \textbf{FLUKA} & \textbf{GENIE} &\textbf{Data}\\
\hline
$\sigma(\nu)$ & $5.3 \cdot 10^{-38} cm^{2}$&$5.2 \cdot 10^{-38} cm^{2}$ &\\
\hline
$CC^{frac}(\nu)$ & $77\%$ & $ 75\%$ &\\
\hline
$\sigma^{CC}(\nu)$ & $4.1\cdot 10^{-38} cm^2$ & $3.9\cdot 10^{-38} cm^2$ &
$4 \cdot 10^{-38} cm^2$
\\
\hline
$\sigma(\overline{\nu})$ & $2.2 \cdot 10^{-38} cm^{2}$&$2.1 \cdot 10^{-38} cm^{2}$  &\\
\hline
$CC^{frac}(\overline{\nu})$ & $ 70\%$ & $ 70\%$ & \\
\hline
$\sigma^{CC}(\overline{\nu})$ & $1.6\cdot 10^{-38} cm^2$ & $1.5\cdot 10^{-38} cm^2$ & $1.6\cdot 10^{-38} cm^2$ \\
\hline
\end{tabular}
\caption{\label{tab:comparison} Comparison between total cross-section per nucleon and fraction of CC events for 5-GeV $\nu_{\mu}$ and $\overline{\nu}_{\mu}$ on a $^{133}$Cs target.}
\end{table}

\begin{figure}[ht]
\centering
\includegraphics[width=.45\textwidth]{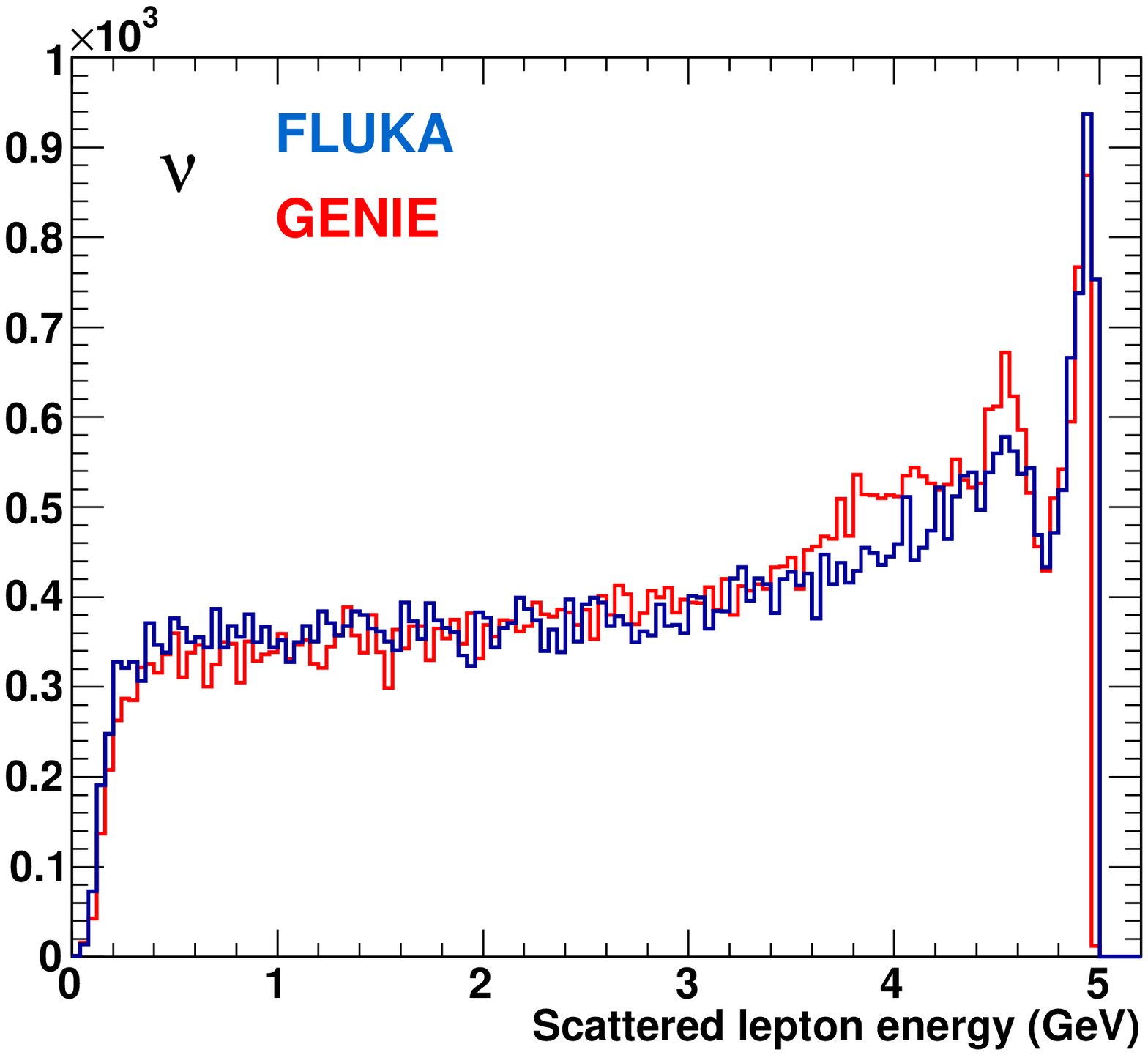}
\includegraphics[width=.45\textwidth]{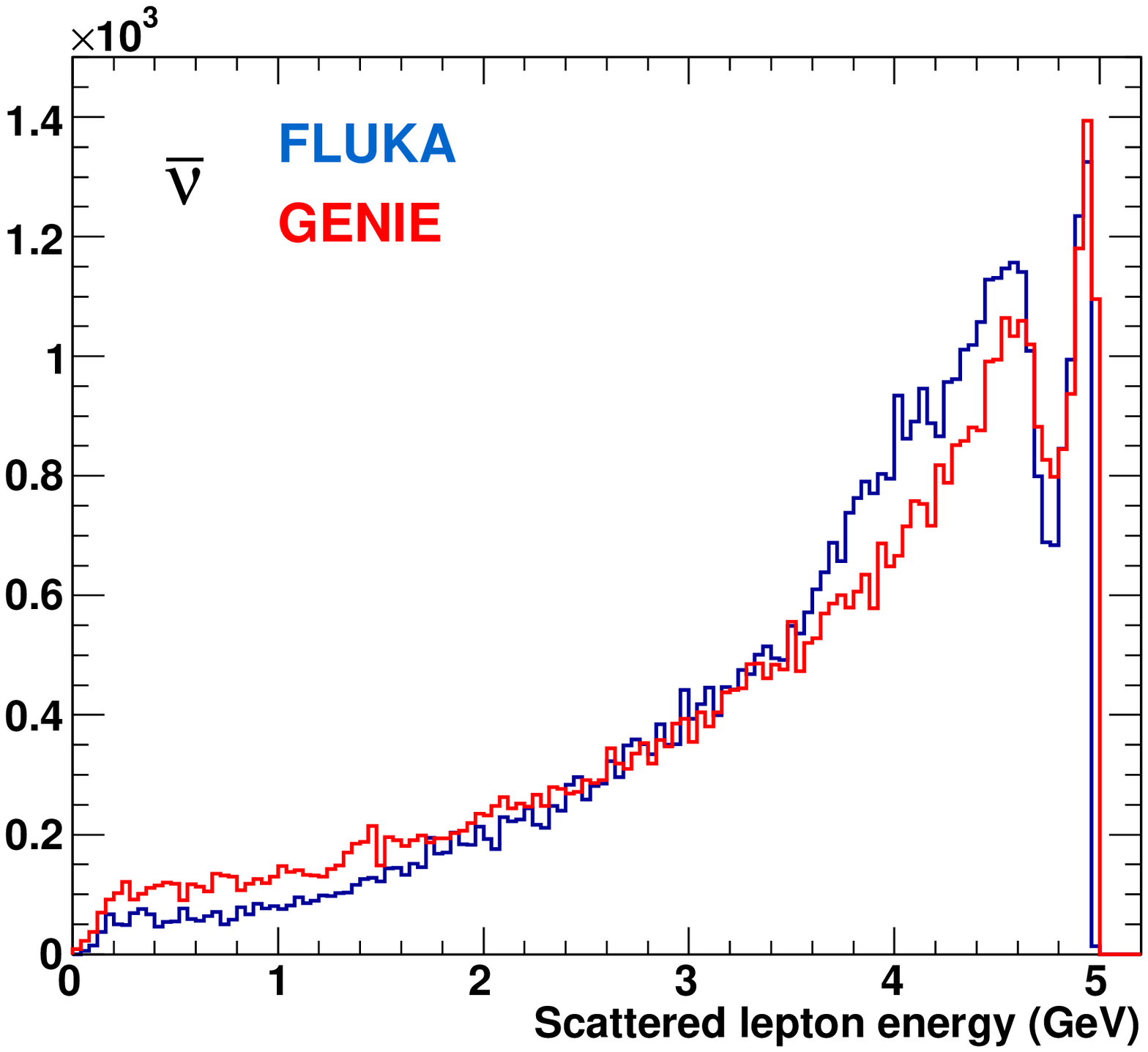}
\caption{\label{fig:comparisonGENIE} Distribution of scattered lepton energy in the reaction $\nu(\overline{\nu})+^{133}Cs \rightarrow l+X$. $l$ can be a neutrino (NC events) or a charged lepton (CC events). All distributions are normalized to the same number of events.}
\end{figure}

Figure~\ref{fig:comparisonGENIE} shows the distribution of the scattered lepton energy (including both CC and NC events), for the case of impinging $\nu$ (left) and impinging $\overline{\nu}$ (right). In both cases, the agreement between FLUKA and GENIE is reasonably good, with a slightly larger discrepancy in case of $\overline{\nu}$. \textit{This confirms the validity of the BDX $\nu$-induced background calculations,} in particular for the most critical background reaction ($\nu_e$ and $\overline{\nu}_e$ CC events, with an high-energy final-state $e^+/e^-$ mimicking the $\chi$ signal.

\begin{figure}[ht]
\centering
\includegraphics[width=.45\textwidth]{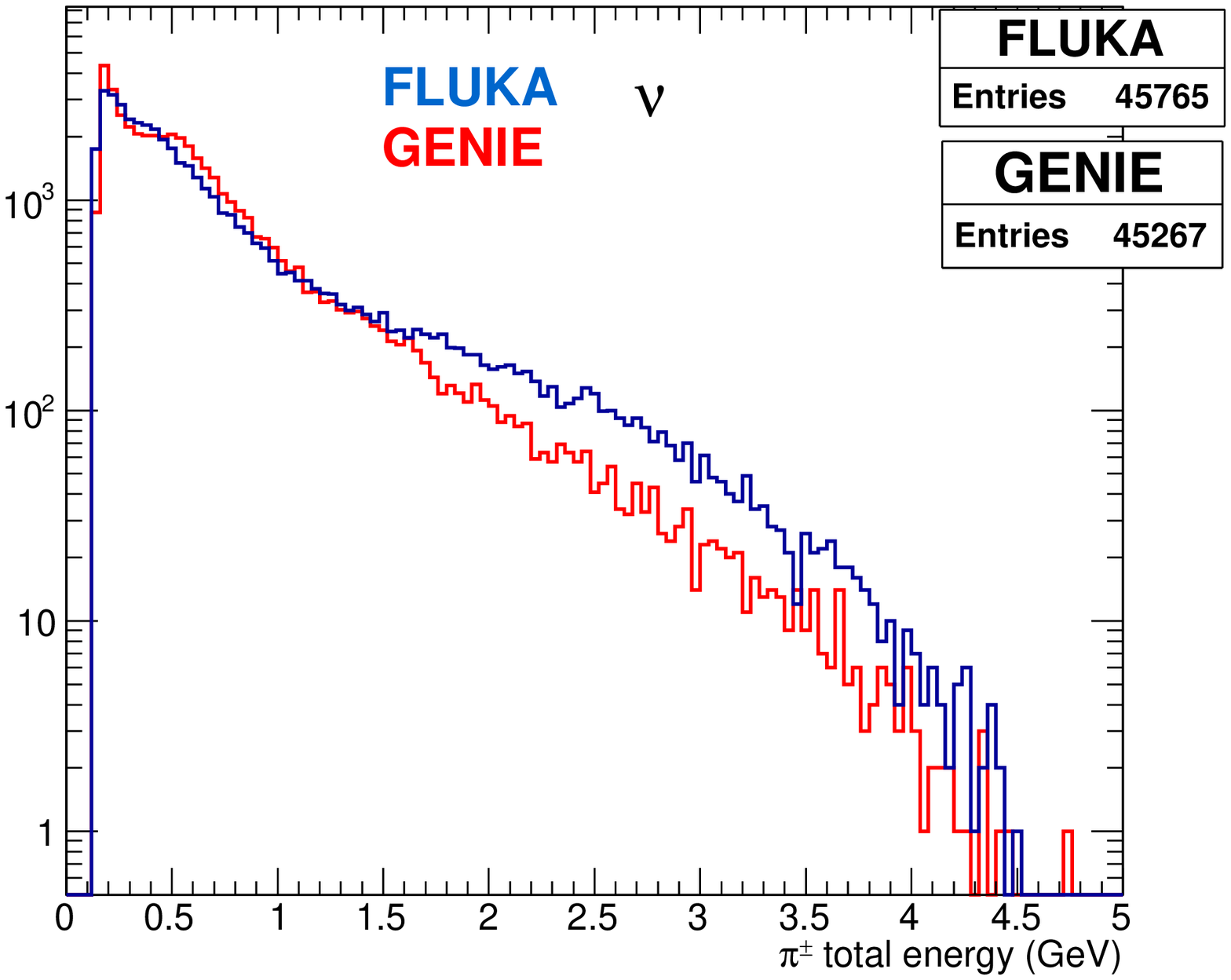}
\includegraphics[width=.45\textwidth]{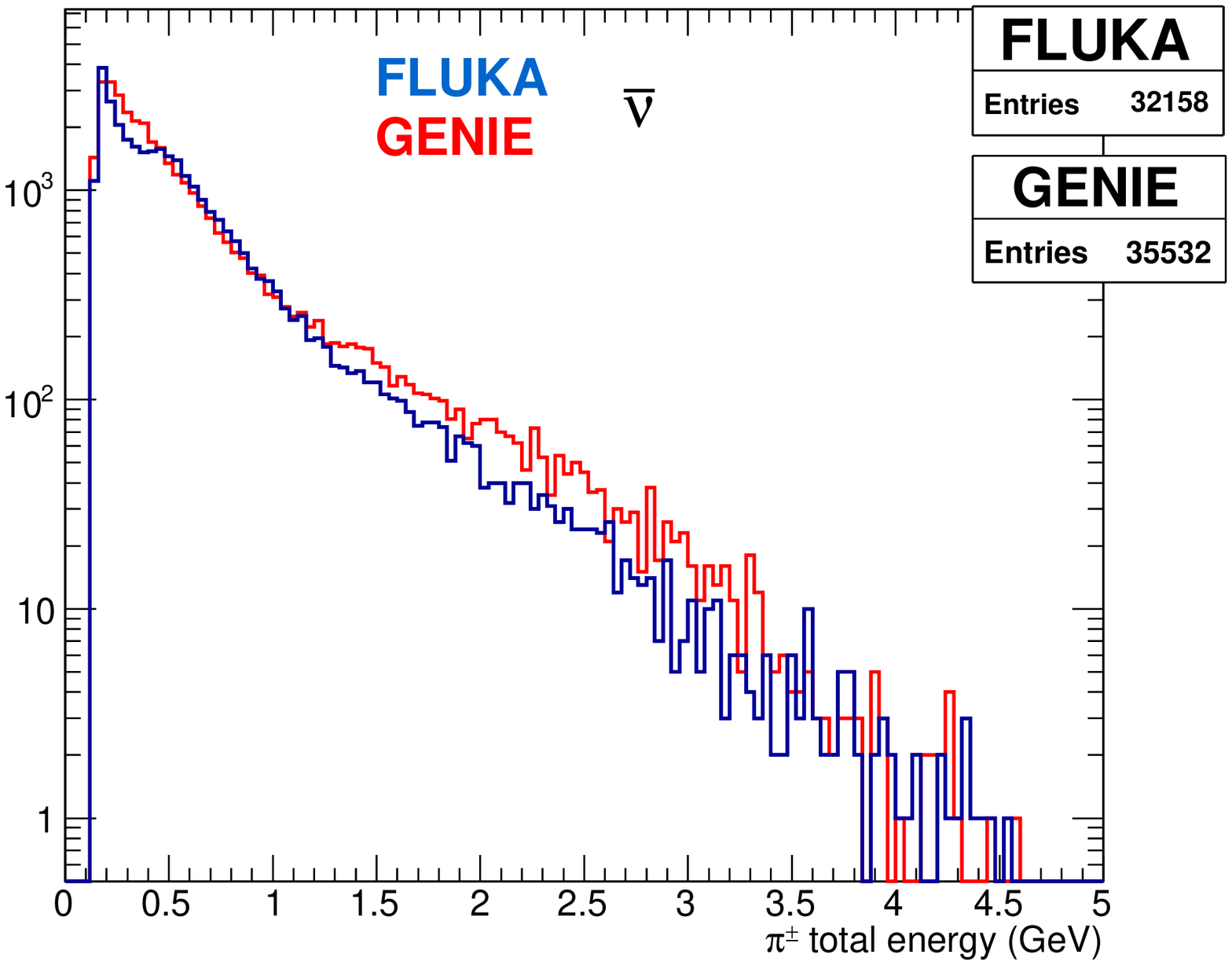}
\caption{\label{fig:comparisonGENIE2} Energy distribution of charged pions in the reaction $\nu(\overline{\nu})+^{133}Cs \rightarrow l+X$. $l$ can be a neutrino (NC events) or a charged lepton (CC events). The two plots refer to the same number of events.}
\end{figure}

\begin{figure}[ht]
\centering
\includegraphics[width=.45\textwidth]{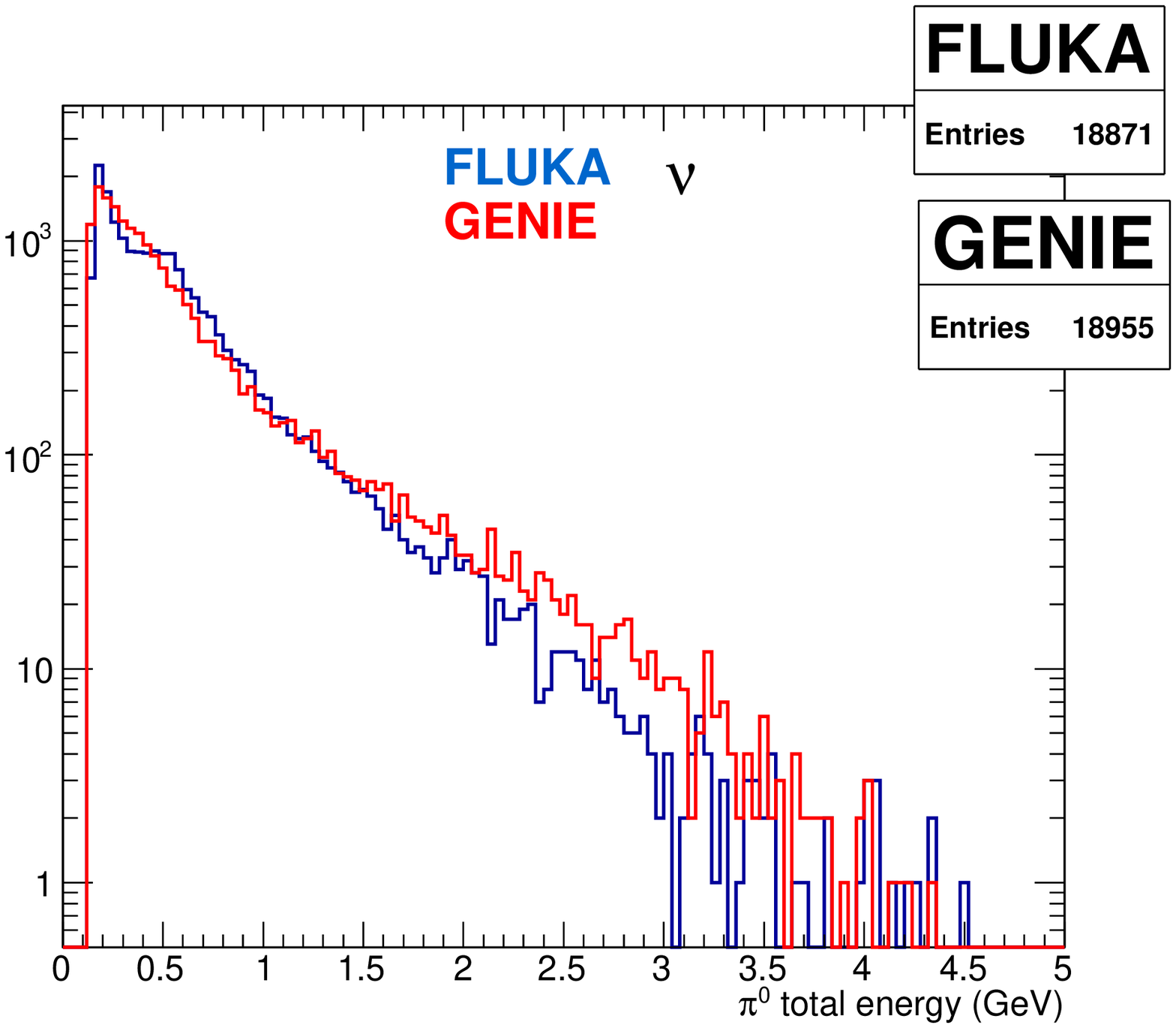}
\includegraphics[width=.45\textwidth]{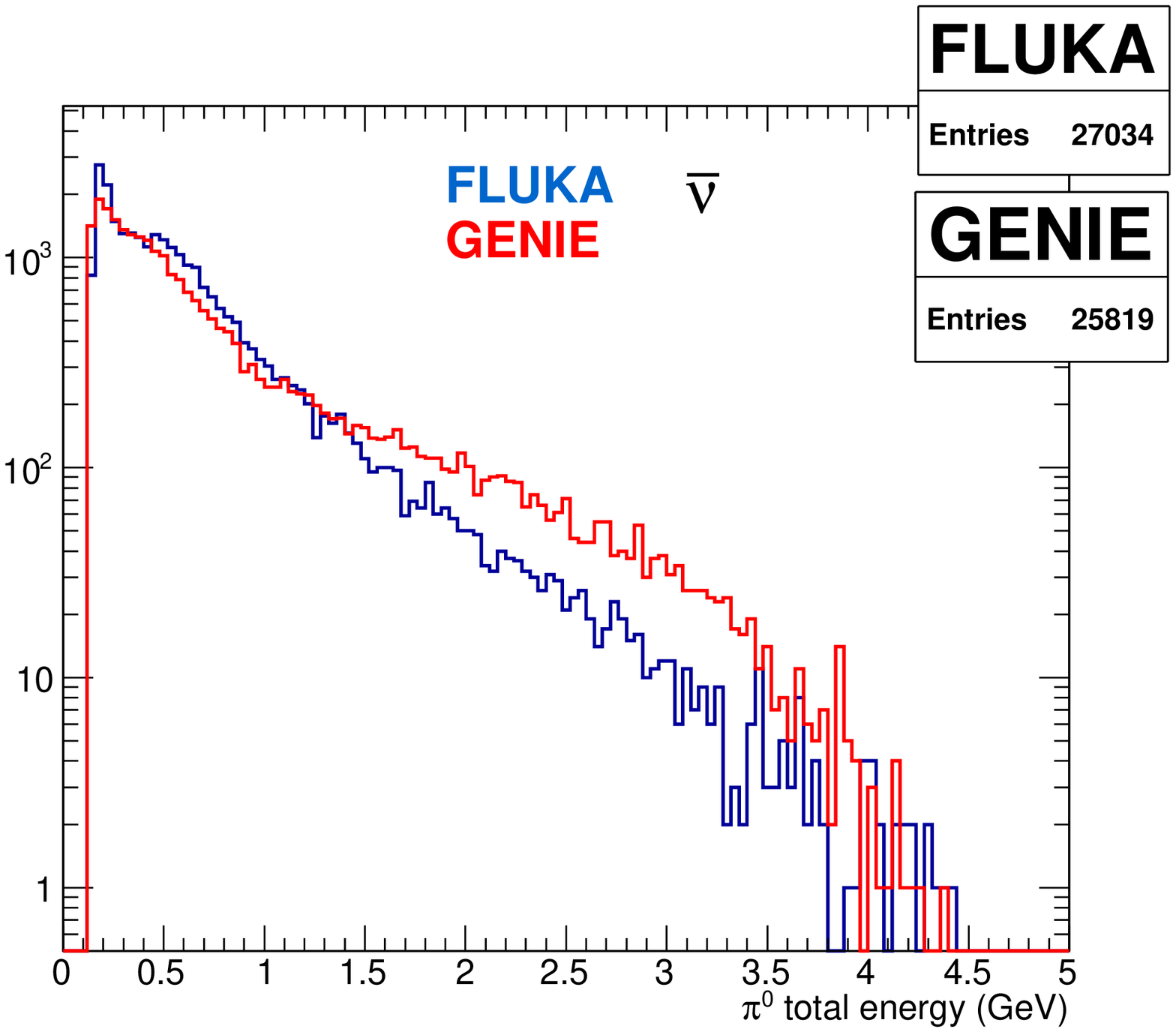}
\caption{\label{fig:comparisonGENIE3} Energy distribution of neutral pions in the reaction $\nu(\overline{\nu})+^{133}Cs \rightarrow l+X$. $l$ can be a neutrino (NC events) or a charged lepton (CC events). The two plots refer to the same number of events.}
\end{figure}

Figure~\ref{fig:comparisonGENIE2} shows the energy distribution of charged pions, for the case of impinging $\nu$ (left) and impinging $\overline{\nu}$ (right). Both distribution refer to the same number of generated events (50k), showing both a good agreement in terms of average multiplicity and shape of the energy distribution. The same conclusion holds for neutral pions (see Fig~\ref{fig:comparisonGENIE3}), a part from a slightly enhanced tail at high energy in GENIE $\overline{\nu}$ results.

%% file: BDX-PAC46-upd-appxC.tex
\section{Comparison of neutrino fluxes with different Hall-A beamline configurations}\label{appx:moller}

In this Appendix, we discuss the neutrino fluxes obtained from the FLUKA simulation of different Hall-A beamline configurations. In particular, we compare the ``simplified'' beam-line configuration, where only the beam-dump was modeled, with other configurations where the following setups were also included:
\begin{itemize}
\item A ``diffuser'' configuration, including the Hall-A diffuser: a $50\%X_0$ Al plate, mounted 17.2 m upstream the beam-dump front-face
\item A ``Moller-like'' configuration, specific to the Moller experimental setup~\cite{moller}, including the 150-cm $LH_s$ target foreseen in the experiment ($\simeq 17\% X_0$), mounted $\simeq$ 50 m upstream the beam-dump front-face.
\end{itemize}
We note that, as concluded from the discussion with Hall-A beamline experts~\cite{welch}, the 150-cm $LH_2$ target foreseen in the experiment ($\simeq 17\% X_0$) provides more than enough beam diffusion to meet the power density goals for the beam-dump, thus allowing to run without the beam-diffuser in place\footnote{The effect of the Moller target on the foreseen signal yield was already discussed in~\cite{moller2}, finding a negligible effect.}. Therefore, in the ``Moller-like'' configuration the diffuser was not included.

\begin{figure}[h]
\centering
\includegraphics[width=0.48\textwidth]{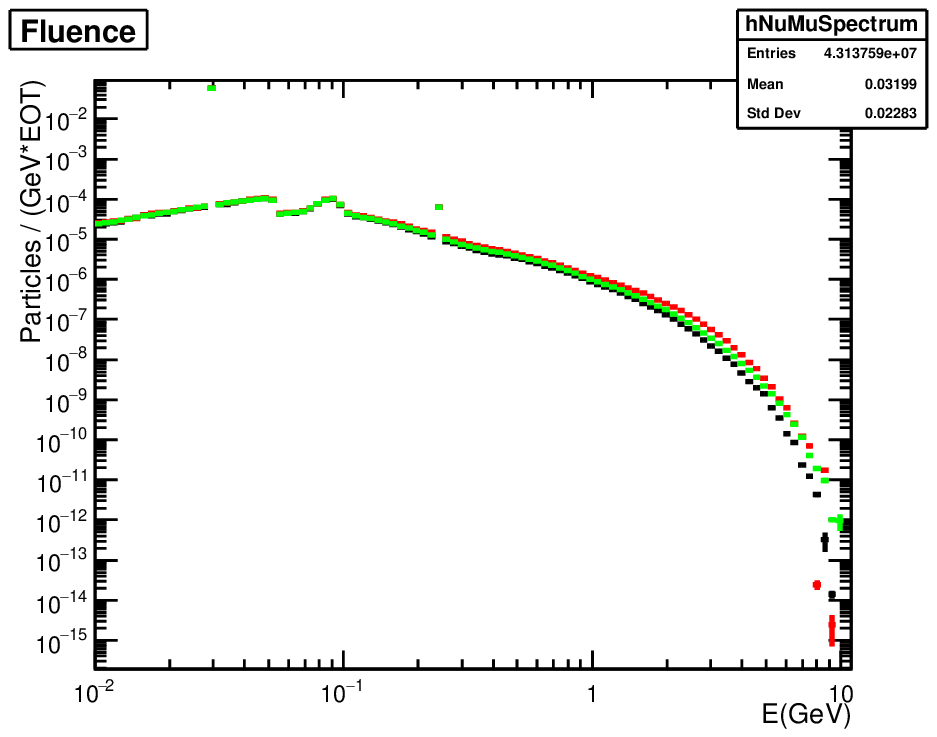}
\includegraphics[width=0.48\textwidth]{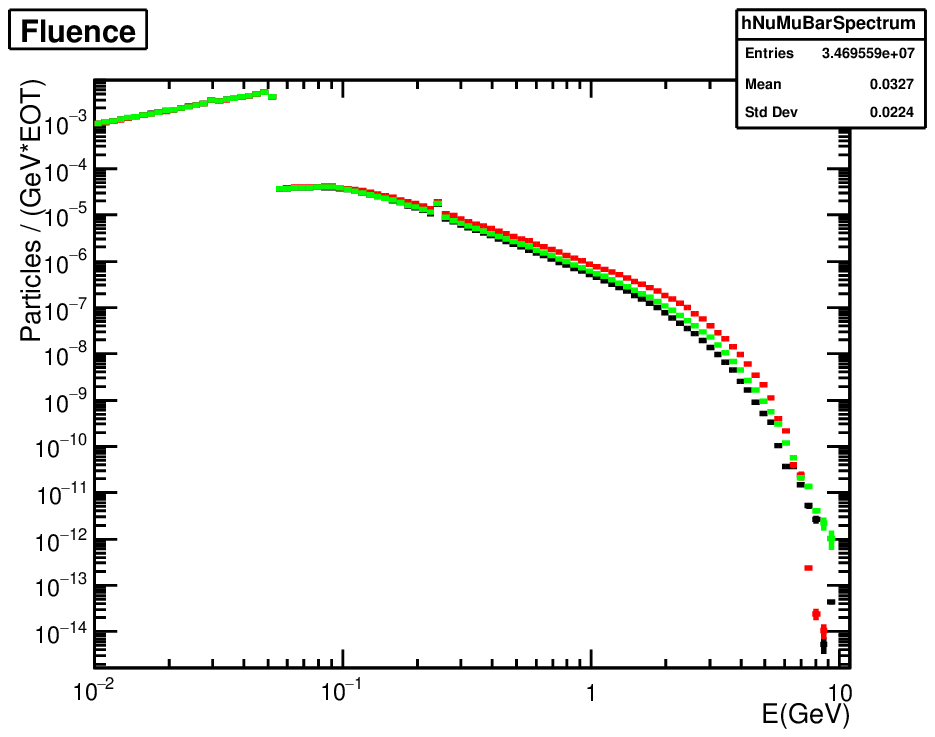}

\includegraphics[width=0.48\textwidth]{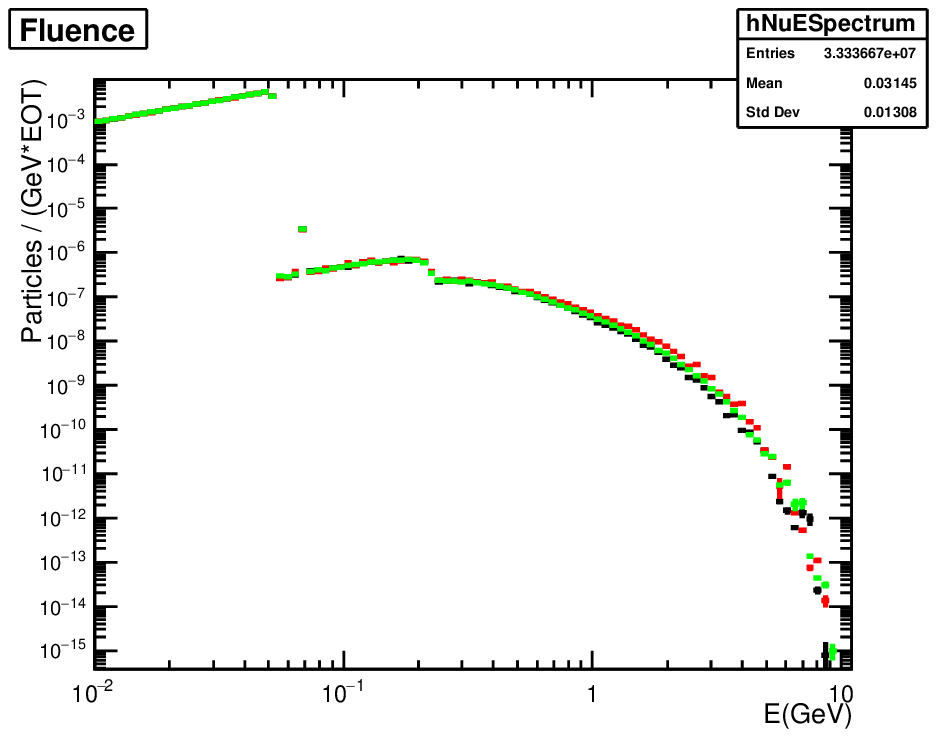}
\includegraphics[width=0.48\textwidth]{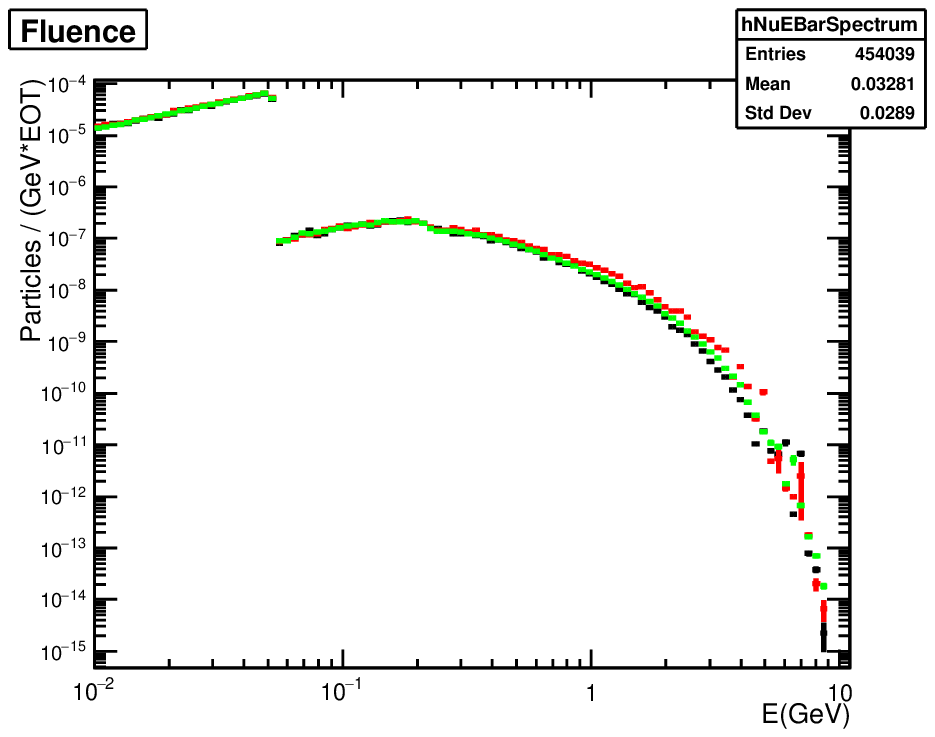}
\caption{\label{fig:fluxcompare} Comparison of $\nu$ fluxes sampled on the front-wall of the BDX experimental Hall, for the ``simplified'' (black), ``diffuser'' (red), and ``Moller-like'' (green) beamline setup. Top-left: $\nu_\mu$, top-right: $\overline{\nu}_\mu$, bottom-left: $\nu_e$, bottom-right: $\overline{\nu}_e$.}
\end{figure}

The neutrino fluxes, sampled on the front wall of the BDX experimental hall, are shown in Fig.~\ref{fig:fluxcompare}, for all neutrino species. Black curves refer to the ``simplified'' beamline configuration, red curves refer to the ``diffuser'' setup, and green curves refer to the ``Moller-like'' configuration.
For all neutrino species, the spectra obtained in the two cases are equivalent at low energy. At higher energies, the configurations including the diffuser or the Moller target result in a slightly larger yield, due to high-energy $\pi$ produced in these beamline elements and decaying in-flight before the beam-dump.

In order to estimate possible effects of different background configurations, we proceeded as follows. The number of background events due to $\nu-N$ interactions in the detector is:
\begin{equation}
N_{\nu-N} \propto \int_{0}^{E_0} dE\, \Phi(E) T(E)\sigma(E) \varepsilon(E)\; \; ,
\end{equation}
where $\Phi(E)$ is the neutrino flux sampled on the BDX experimental hall front-wall, $T(E)$ is the transport function from the front-wall to the detector volume, $\sigma(E)$ is the $\nu-N$ cross-section, and $\varepsilon(E)$ is the detector response efficiency. The energy dependence of above quantities is roughly as follows:
\begin{itemize}
\item $T(E) \simeq 10^{-2}+7\cdot10^{-3}E$: high-energy $\nu$ are typically emitted in the forward direction, thus entering the detector acceptance.
\item $\sigma(E) \propto s=(M_n^2+2M_n E)$, with $M_n$ the nucleon mass. 
\item $\varepsilon(E) \simeq A \cdot (E-B)$, with $A=1.3\% (\nu_\mu)$ / $7\% (\nu_e)$ and $B=0.7$ GeV $(\nu_\mu$) / $0.5$ GeV $(\nu_e)$.
\end{itemize}

\begin{table}[h]
\centering
\begin{tabular}{|c|c|c|}
\hline
& ``diffuser'' & ``Moller''\\
\hline
$\nu_\mu$ & 2.14 & 1.43 \\
\hline
$\nu_e$ &1.75 &1.25 \\
\hline
$\overline{\nu}_\mu$ &2.57 &1.46 \\
\hline
$\overline{\nu}_e$ &1.92 &1.26 \\
\hline
\end{tabular}
\caption{\label{tab:moller} Neutrino-induced background rates, normalized to the ``simplified'' configuration, for the ``diffuser'' and for the ``Moller'' configurations.}
\end{table}

The results of the calculation are reported in Tab.~\ref{tab:moller}, in terms of the ratio between the new configuration (``diffuser'' or ``Moller'') and the ``simplified'' one. The ``diffuser'' configuration foresees a larger neutrino background, of about a factor 2.
Considering that the experiment sensitivity on $\varepsilon^2$ scale as the number of background counts to the 1/4 power, this would result in a $\lesssim 20\%$ worse reach, an effect almost negligible in the logarithmic reach plot. 

This demonstrates that the BDX experiment is fully compatible with both the ``standard'' Hall-A beam-line configuration, using the diffuser with a thin-target, and with the ``Moller'' configuration. At the same time it provides input to the Hall A beam line configuration from the BDX experiment.